\documentclass[aps,prb,twocolumn,superscriptaddress,longbibliography]{revtex4-2}
\usepackage{lineno}
\usepackage{graphicx}  
\usepackage{dcolumn}   
\usepackage{bm}        
\usepackage{amssymb}   
\usepackage{amsmath}
\usepackage{blkarray, multirow, graphicx, diagbox, color, xcolor, colortbl}
\usepackage{bbm, bbold}
\usepackage{glossaries}
\usepackage{hyphenat}
\usepackage{ifthen}
\usepackage{xkeyval}
\usepackage{moreverb}
\usepackage{rotating}
\usepackage{wrapfig}
\usepackage{slashbox}
\usepackage{xspace}
\usepackage{nicefrac}
\usepackage[]{units}
\usepackage{physics}
\usepackage{braket}
\usepackage[inline]{enumitem}
\usepackage{tabto}
\usepackage{listings}
\usepackage{xstring}
\def\ReplaceStr#1{%
	\IfSubStr{#1}{p}{%
		\StrSubstitute{#1}{p}{.}}{#1}}

\usepackage[caption=false]{subfig}
\usepackage[customcolors]{hf-tikz} 
\usepackage{tikz}
\usepackage{calc}
\usepackage{pgffor}
\usepackage{pgfplots}
\pgfplotsset{compat=1.13}
\usepackage{pgfplotstable}
\usepgfplotslibrary{groupplots}
\usepgfplotslibrary{fillbetween}

\tikzstyle{n} = [draw,shape=ellipse,minimum size=1.5em,inner sep=0pt,fill=white!20, minimum width=2.5em]
\tikzstyle{Init} = [n,color=green,fill=green!20,text=black]
\tikzstyle{Fin} = [n,color=red,fill=red!20,text=black]
\tikzstyle{Ghost} = [minimum size=1.5em,inner sep=0pt,color=white,text=black]
\tikzstyle{Multiple} = [draw,shape=rect,minimum size=2em,inner sep=0pt]

\tikzstyle{ghostA} = [text=red!70,thick, minimum size=2*(5pt-\pgflinewidth), inner sep=0pt, outer sep=0pt]
\tikzstyle{ghostB} = [text=blue!70,thick, minimum size=2*(5pt-\pgflinewidth), inner sep=0pt, outer sep=0pt]
\tikzstyle{siteA} = [regular polygon, regular polygon sides=3, shape border rotate= 30, draw=red!50,fill=red!20,thick,inner sep=0pt,minimum width=1.5em,font=\footnotesize]
\tikzstyle{siteB} = [regular polygon, regular polygon sides=3, shape border rotate= -30, draw=green!50,fill=green!20,thick,inner sep=0pt,minimum width=1.5em,font=\footnotesize]
\tikzstyle{op} = [regular polygon, regular polygon sides=4, draw=orange!50, fill=orange!20, thick, inner sep=0.2pt, minimum width=1.25em, minimum height=1.5em,font=\footnotesize]
\tikzstyle{opghost} = [regular polygon, regular polygon sides=4, thick, inner sep=0.2pt, minimum width=1.25em, minimum height=1.5em,font=\footnotesize]
\tikzstyle{site} = [circle,draw=blue!50,fill=blue!20,thick,inner sep=0.2pt,minimum width=1.25em,font=\footnotesize]
\tikzstyle{hiddensite} = [circle,draw=white!50,fill=white!20,thick,inner sep=0.2pt,minimum width=1.25em,font=\footnotesize]
\tikzstyle{nosite} = [circle,draw=white,fill=white,thick,inner sep=0.1pt,minimum width=1.5em]
\tikzstyle{ghost} = [font=\footnotesize]
\tikzstyle{intersite} = [regular polygon, regular polygon sides=4, shape border rotate= 45, draw=black!50,fill=black!20,thick,inner sep=0pt,minimum width=1.5em]
\tikzstyle{ld} = [inner sep=1pt, font=\small]
\tikzstyle{unsite} = [circle, outer sep=0pt,inner sep=0.2pt,minimum width=1.25em]

\definecolor{colorA}{rgb} {0.58,0,0.8275}
\definecolor{colorB}{rgb} {0.11,0.663,0.51}
\definecolor{colorC}{rgb} {0.3373,0.7059,0.9137}
\definecolor{colorD}{rgb} {0.902,0.6235,0}
\definecolor{colorE}{rgb} {0.9451,0.902,0.3255}
\definecolor{colorF}{rgb} {0.3373,0.3255,0.902}
\definecolor{colorG}{rgb} {0.9451,0.3255,0.3373}

\usetikzlibrary
{
	calc,
	decorations,
	plotmarks,
	patterns,
	positioning,
	petri,
	arrows,
	decorations.markings,
	backgrounds,
	fit,
	graphs,
	shapes.geometric,
	decorations.pathmorphing,
	shapes.misc,
	shapes,
	tikzmark,
	pgfplots.colorbrewer,
	fpu,
}

\usepackage{ocgx}

\usetikzlibrary{ocgx}

\pgfplotsset{
        cycle from colormap manual style/.style={
            x=3cm,y=10pt,ytick=\empty,
            stack plots=y,
            every axis plot/.style={line width=2pt},
        },
}

\tikzset{->-/.style={decoration={
			markings,
			mark=at position .5 with {\arrow{>}}},postaction={decorate}}}

\tikzset{-<-/.style={decoration={
			markings,
			mark=at position .5 with {\arrow{<}}},postaction={decorate}}}

\tikzstyle{orientedsnake} = [
decorate, 
decoration={snake},
->
]  
\tikzstyle{orientedshortarrow} = [
decoration={markings,
	mark=at position .33 with {\arrow{>}}},
postaction={decorate}
]  
\tikzstyle{orientedlongarrow} = [
decoration={markings,
	mark=at position .67 with {\arrow{>}}},
postaction={decorate}
]
\tikzset{dbl/.style={double,
		double equal sign distance,
		-implies,
		shorten >=10pt,
		shorten <=10pt}}
\tikzset{
	between/.style args={#1 and #2}{
		at = ($(#1)!0.5!(#2)$)
	}
}

\newcommand{\nodagger}[0]{{\phantom{\dagger}}}

\pgfmathdeclarefunction{peierlspotential}{6}%
{
	\pgfmathparse
	{
		#2 / (#3 * #5) 
		* sin(deg(#1 * #3 - #3 * #5 * (x)))
		* exp(-( ( #1 - #5 * ((x) - #4) ) )^2.0 / ( #6^2.0 ) )
	}%
}

\pgfmathdeclarefunction{linearFct}{2}%
{%
	\pgfmathparse{#1*x+#2}%
}
\pgfmathdeclarefunction{quadFct}{3}%
{%
	\pgfmathparse{#1*x*x+#2*x+#3}%
}
\pgfmathdeclarefunction{logFct}{2}%
{%
	\pgfmathparse{#1*log10(x)+#2}%
}
\pgfmathdeclarefunction{expFct}{2}%
{%
	\pgfmathparse{#1*exp(-x/#2)}%
}

\newboolean{buildCurrentBlockInline}
\setboolean{buildCurrentBlockInline}{false}
\newboolean{rebuildCurrentBlock}
\setboolean{rebuildCurrentBlock}{false}

\newboolean{rebuildData}
\setboolean{rebuildData}{false}

\newboolean{removeData}
\setboolean{removeData}{false}

\newboolean{buildtikzpics}
\setboolean{buildtikzpics}{true}
\newif\ifrebuildtikz
\newif\ifChangeMode
\ChangeModetrue
\ChangeModefalse
\ifthenelse{\boolean{buildtikzpics}}
{
	\rebuildtikztrue
	\usetikzlibrary{external}
	\tikzexternalize[optimize=false,prefix=Figures/autogen/]
}
{
	\rebuildtikzfalse
}
\usepackage[colorlinks,bookmarks=false,citecolor=blue,linkcolor=black,urlcolor=blue]{hyperref}
\usepackage{cleveref}
\Crefname{appendix}{Appendix}{Appendices}
\Crefname{equation}{Equation}{Equations}
\Crefname{figure}{Figure}{Figures}
\Crefname{section}{Section}{Sections}
\Crefname{tabular}{Tabular}{Tabulars}
\crefname{appendix}{App.}{Apps.}
\crefname{equation}{Eq.}{Eqs.}
\crefname{figure}{Fig.}{Figs.}
\crefname{section}{Sec.}{Secs.}
\crefname{tabular}{Tab.}{Tabs.}

\ExplSyntaxOn
\DeclareExpandableDocumentCommand \eval { m } { \fp_eval:n { #1 } }
\ExplSyntaxOff

\makeatletter
\def\pgfplotsutil@decstringcounter#1{%
 \begingroup
  \c@pgf@counta=#1\relax
  \advance\c@pgf@counta by -1
  \edef#1{\the\c@pgf@counta}%
  \pgfmath@smuggleone#1%
 \endgroup
}%

\pgfplotsset{
/pgfplots/each nth point*/.style 2 args={%
/pgfplots/x filter/.append code={%
 \ifnum\coordindex=0
  \def\c@pgfplots@eachnthpoint@xfilter{0}%
  \edef\c@pgfplots@eachnthpoint@xfilter@cmp{#1}%
 \else
  \ifnum\coordindex>#2\relax
   \pgfplotsutil@advancestringcounter\c@pgfplots@eachnthpoint@xfilter
   \ifx\c@pgfplots@eachnthpoint@xfilter@cmp\c@pgfplots@eachnthpoint@xfilter
    \def\c@pgfplots@eachnthpoint@xfilter{0}%
   \else
    \let\pgfmathresult\pgfutil@empty
   \fi
  \fi
 \fi
}%
},
}
\makeatother
\usepackage{mathtools}
\usepackage{braket}
\usepackage{amsmath, amssymb}
\usepackage{graphicx}
\usepackage{dcolumn}
\usepackage{bm}
\usepackage{float}
\usepackage{csvsimple}
\usepackage{nicefrac}
\usepackage{glossaries}
\usepackage{placeins}
\graphicspath{{Figures/}}
\usepackage{tikz}
\usepackage{soul}
\usepackage{xcolor}
\newacronym[shortplural={MPS}]{MPS}{MPS}{matrix\hyp product state}
\newacronym{MF}{MF}{mean\hyp field}
\newacronym{BCS}{BCS}{Bardeen\hyp Cooper\hyp Schrieffer}
\newacronym{ACA}{ACA}{adaptive cross approximation}
\newacronym{1D}{1D}{one\hyp dimensional}
\newacronym{2D}{2D}{two\hyp dimensional}
\newacronym{MPO}{MPO}{matrix\hyp product operator}
\newacronym{SVD}{SVD}{singular\hyp value decomposition}
\newacronym{QCS}{QCS}{quantum\hyp computer simulator}
\newacronym{QC}{QC}{quantum computer}
\newacronym{FSM}{FSM}{finite\hyp state machine}
\newacronym{CDW}{CDW}{charge\hyp density wave}
\newacronym{SDW}{SDW}{spin\hyp density wave}
\newacronym{ARPES}{ARPES}{angle-resolved photoemission spectroscopy}
\newacronym{OBC}{OBC}{open-boundary conditions}
\newacronym{PBC}{PBC}{periodic-boundary conditions}
\newacronym{TEBD}{TEBD}{time-evolution block-decimation}
\newacronym{TDVP}{TDVP}{time\hyp dependent variational principle}
\newacronym{iff}{iff}{if and only if}
\newacronym{DFT}{DFT}{density\hyp functional theory}
\newacronym{DMFT}{DMFT}{dynamical mean\hyp field theory}
\newacronym{DMRG}{DMRG}{density\hyp matrix renormalization group}
\newacronym{1DMRG}{1DMRG}{single-site density\hyp matrix renormalization group}
\newacronym{2DMRG}{2DMRG}{two-site density\hyp matrix renormalization group}
\newacronym{DMRG3S}{DMRG3S}{strictly single-site density\hyp matrix renormalization group}
\newacronym{iDMRG}{iDMRG}{inifinite\hyp size density\hyp matrix renormalization group}
\newacronym{tDMRG}{tDMRG}{time\hyp dependent density\hyp matrix renormalization group}
\newacronym{PP-DMRG}{PP-DMRG}{projected purified density\hyp matrix renormalization group}
\newacronym{QMC}{QMC}{quantum Monte Carlo}
\newacronym{AIM}{AIM}{Anderson impurity model}
\newacronym{SIAM}{SIAM}{single impurity Anderson model}
\newacronym{LDA}{LDA}{local\hyp density approximation}
\newacronym{VQE}{VQE}{variational\hyp quantum eigensolver}
\newacronym{ED}{ED}{exact diagonalization}
\newacronym{QPT}{QPT}{quantum phase transition}
\newacronym{QCP}{QCP}{quantum critical point}
\newacronym{ETH}{ETH}{eigenstate thermalization hypothesis}
\newacronym{EHM}{EHM}{extended Hubbard model}
\newacronym{AKLT}{AKLT}{Affleck\hyp Lieb\hyp Kennedy\hyp Tasaki}
\newglossaryentry{QR}{name={QR},description={QR decomposition}}
\newacronym{TNS}{TNS}{tensor\hyp network state}
\newacronym{SM}{SM}{supplemental material}
\newacronym{NOO}{NOO}{natural orbital occupation}
\newacronym{NO}{NO}{natural orbital}
\newacronym{LRO}{LRO}{long\hyp range order}
\newacronym{qLRO}{qLRO}{quasi\hyp long\hyp range order}
\newacronym{SC}{SC}{superconductivity}
\newacronym{VBGS}{VBGS}{valence bond ground-state}
\newacronym{PEPS}{PEPS}{projected entangled pair\hyp states}
\newacronym{ALS}{ALS}{alternating least squares}
\newacronym{BdG}{BdG}{Bogoljubov de-Gennes}
\newacronym{TFIM}{TFI}{transverse field Ising model}
\newacronym{PP}{PP}{projected purification}
\newacronym{BEC}{BEC}{Bose\hyp Einstein condensate}
\newacronym{JWT}{JWT}{Jordan\hyp Wigner transformation}
\newacronym{NISQ}{NISQ}{noisy intermediate scale quantum}
\newacronym{NN}{NN}{nearest\hyp neighbor}
\newacronym{NNN}{NNN}{next\hyp nearest\hyp neighbor}
\newacronym{SPDM}{SPDM}{single\hyp particle density matrix} 
\newacronym{HCB}{HCB}{hardcore bosons}
\newacronym{SF}{SF}{spinless fermions}
\newacronym{fRG}{fRG}{functional renormalization group}
\newacronym{LE}{LE}{Luther\hyp Emery}
\newacronym{TLL}{TLL}{Tomonaga-Luttinger liquid}
\newacronym{PLL}{PLL}{pair Luttinger liquid}
\newacronym{FB}{FB}{flat band}
\newacronym{SU}{SU}{superfluid}
\newacronym{QO}{QO}{quasi-order}
\newacronym{LL}{LL}{Luttinger liquid}
\newacronym{CFT}{CFT}{conformal field theory}
\newacronym{USC}{USC}{unconventional superconductivity}
\newacronym{Q1D}{Q1D}{quasi-one-dimensional}
\usetikzlibrary{external,arrows}
\newcommand{\Systems}[7]%
{%
	\pgfmathparse{int((#2*#5-1))}%
	\pgfmathsetmacro{\ymax}{\pgfmathresult}%
	\pgfmathparse{int(#1-1)}%
	\pgfmathsetmacro{\xmax}{\pgfmathresult}%
	\tikzset{external/export next=false}%
	\begin{tikzpicture}[baseline]%
		\foreach \x in {0,...,\xmax}%
		{%
			\foreach \y in {0,...,\ymax}%
			{%
				\node[circle, draw, inner sep = 0pt, outer sep = 0pt] at ($(\x*1.5*#3,\y*1.25*#3)$) (\x\y){\pgfmathparse{int(10*#3)}\fontsize{\pgfmathresult pt}{10pt}#4};%
			}%
		}%
		\foreach \x [remember=\x as \lastx (initially 0)] in {0,...,\xmax}%
		{%
			\foreach \y [remember=\y as \lasty (initially 1)] in {0,...,\ymax}%
			{%
				\ifthenelse{\y>0}%
				{%
					\pgfmathparse{int(mod(\y,#2))}%
					\ifthenelse{\pgfmathresult=0}%
					{%
						\draw[densely dotted] (\x\lasty) -- node[left, inner ysep = 0.5em] {} (\x\y);%
					}%
					{%
						\draw (\x\lasty) -- node[left, inner ysep = 0.5em] {} (\x\y);%
					}%
				}{}%
				\ifthenelse{\ymax>0}%
				{%
					\draw[densely dotted, opacity=#7] (\lastx\lasty) -- node[midway] (m) {} (\x\y);%
				}%
				{}%
				\ifthenelse{\x>0}%
				{%
					\pgfmathparse{int(mod(\x,#2))}%
					\ifthenelse{\pgfmathresult=0}%
					{%
						\begin{scope}[on background layer]
							\draw[opacity=#6] (\lastx\y.east) -- node[above, inner xsep = 0.5em, inner ysep = 0.1em] {} (\x\y.west);%
						\end{scope}
					}%
					{%
						\draw (\lastx\y) -- node[above, inner xsep = 0.5em, inner ysep = 0.1em] {} (\x\y);%
					}%
				}{}%
			}%
		}%
		\pgfmathparse{int((#5-1)/2*(#2))}%
		\pgfmathsetmacro{\yPhysMin}{\pgfmathresult}%
		\pgfmathparse{int(\yPhysMin+#2-1)}%
		\pgfmathsetmacro{\yPhysMax}{\pgfmathresult}%
		\pgfmathparse{int(#1-1)}%
		\pgfmathsetmacro{\xmax}{\pgfmathresult}%
		\node[draw, fit = (0\yPhysMin) (\xmax\yPhysMax)] (Phys) {};
		\node[anchor=west] at (Phys.north east) {$H_0$};
	\end{tikzpicture}%
}%
\newcommand{\Fermions}[5]%
{%
	\pgfmathparse{rnd}%
	\pgfmathsetmacro{\uprnd}{\pgfmathresult}%
	\ifthenelse{\lengthtest{\uprnd pt > #1pt }}%
	{%
		{\color{#2}$\uparrow$}%
	}%
	{%
		\color{#3}$\uparrow$%
	}%
	\pgfmathparse{rnd}%
	\pgfmathsetmacro{\downrnd}{\pgfmathresult}%
	\ifthenelse{\lengthtest{\downrnd pt > #1pt }}%
	{%
		{\color{#4}$\downarrow$}%
	}%
	{%
		{\color{#5}$\downarrow$}%
	}%
}%
\newcommand{\Bosons}[3]%
{%
	\pgfmathparse{rnd}%
	\pgfmathsetmacro{\rnd}{\pgfmathresult}%
	\ifthenelse{\lengthtest{\rnd pt > #1pt }}%
	{%
		{\color{#2}\textbullet}%
	}%
	{%
		{\color{#3}\textbullet}%
	}%
}%

\definecolor{colorA}{rgb} {0.58,0,0.8275}
\definecolor{colorB}{rgb} {0.11,0.663,0.51}
\definecolor{colorC}{rgb} {0.3373,0.7059,0.9137}
\definecolor{colorD}{rgb} {0.902,0.6235,0}
\definecolor{colorE}{rgb} {0.9451,0.902,0.3255}
\definecolor{colorF}{rgb} {0.3373,0.3255,0.902}
\definecolor{colorG}{rgb} {0.9451,0.3255,0.3373}
\definecolor{colorH}{rgb} {0.11,0.3255,0.3373}

\pgfmathdeclarefunction{critExponentFct}{3}%
{%
	\pgfmathparse{(x<#2)?(#1*abs(x-#2)^(#3)):(0)}%
}

\pgfmathdeclarefunction{exponentialInFct}{3}%
{%
	\pgfmathparse{#1+#2*exp(-(#3)/x)}%
}

\newcommand{\printpgfnumberwithouterrorInMath}[3][0]%
{%
	\pgfmathfloatparsenumber{#2}%
	\pgfmathfloattomacro{\pgfmathresult}{\Fn}{\Mn}{\En}%
	\pgfmathparse{#1==2 ? (\Fn==2 ? "+" : "-") : (\Fn==2 ? "-" : (#1==1 ? "+" : ""))}%
	\edef\Sn{\pgfmathresult}%
	\pgfmathfloatparsenumber{#3}%
	\pgfmathfloattomacro{\pgfmathresult}{\Fe}{\Me}{\Ee}%
	\pgfmathparse{int(sqrt((\Ee-\En)^2))}%
	\edef\precisionAbsEe{\pgfmathresult}%
	\pgfmathparse{int(\Ee-\En)}%
	\edef\precisionE{\pgfmathresult}%
	\pgfmathparse{\eval{\Me*10^(\precisionE)}}%
	\ifthenelse{\En=0}%
	{%
		\Sn\pgfmathprintnumber[fixed, precision=\precisionAbsEe, zerofill]{\Mn}%
	}%
	{%
		\ifthenelse{\En=1}%
		{%
			\pgfmathparse{\eval{\Mn*10}}%
			\edef\Mn{\pgfmathresult}%
			\pgfmathparse{int(\precisionAbsEe+1)}%
			\edef\precisionAbsEe{\pgfmathresult}%
			\Sn\pgfmathprintnumber[fixed, precision=\precisionAbsEe, zerofill]{\Mn}%
		}%
		{%
			\ifthenelse{\En=2}%
			{%
				\pgfmathparse{\eval{\Mn*100}}%
				\edef\Mn{\pgfmathresult}%
				\pgfmathparse{int(\precisionAbsEe+1)}%
				\edef\precisionAbsEe{\pgfmathresult}%
				\Sn\pgfmathprintnumber[fixed, precision=\precisionAbsEe, zerofill]{\Mn}%
			}%
			{%
				\ifthenelse{\En=-1}%
				{%
					\pgfmathparse{\eval{\Mn/10}}%
					\edef\Mn{\pgfmathresult}%
					\pgfmathparse{int(\precisionAbsEe+1)}%
					\edef\precisionAbsEe{\pgfmathresult}%
					\Sn\pgfmathprintnumber[fixed, precision=\precisionAbsEe, zerofill]{\Mn}%
				}%
				{%
					\ifthenelse{\En=-2}%
					{%
						\pgfmathparse{\eval{\Mn/10}}%
						\edef\Mn{\pgfmathresult}%
						\pgfmathparse{int(\precisionAbsEe+1)}%
						\edef\precisionAbsEe{\pgfmathresult}%
						\Sn\pgfmathprintnumber[fixed, precision=\precisionAbsEe, zerofill]{\Mn}%
					}%
					{%
						\Sn\pgfmathprintnumber[std, precision=\precisionAbsEe, zerofill]{\Mn} \cdot10^{\En}%
					}%
				}%
			}%
		}%
	}%
}

\tikzstyle{ghostA} = [text=red!70,thick, minimum size=2*(5pt-\pgflinewidth), inner sep=0pt, outer sep=0pt]
\tikzstyle{ghostB} = [text=blue!70,thick, minimum size=2*(5pt-\pgflinewidth), inner sep=0pt, outer sep=0pt]
\tikzstyle{siteA} = [regular polygon, regular polygon sides=3, shape border rotate= 30, draw=red!50,fill=red!20,thick,inner sep=0pt,minimum width=1.5em,font=\footnotesize]
\tikzstyle{siteB} = [regular polygon, regular polygon sides=3, shape border rotate= -30, draw=green!50,fill=green!20,thick,inner sep=0pt,minimum width=1.5em,font=\footnotesize]
\tikzstyle{op} = [regular polygon, regular polygon sides=4, draw=orange!50, fill=orange!20, thick, inner sep=0.2pt, minimum width=1.25em, minimum height=1.5em,font=\footnotesize]
\tikzstyle{opghost} = [regular polygon, regular polygon sides=4, thick, inner sep=0.2pt, minimum width=1.25em, minimum height=1.5em,font=\footnotesize]
\tikzstyle{site} = [circle,draw=blue!50,fill=blue!20,thick,inner sep=0.2pt,minimum width=0.75em,font=\footnotesize]
\tikzstyle{hiddensite} = [circle,draw=white!50,fill=white!20,thick,inner sep=0.2pt,minimum width=1.25em,font=\footnotesize]
\tikzstyle{nosite} = [circle,draw=white,fill=white,thick,inner sep=0.2pt,minimum width=1.25em]
\tikzstyle{ghost} = [font=\footnotesize]
\tikzstyle{intersite} = [regular polygon, regular polygon sides=4, shape border rotate= 45, draw=black!50,fill=black!20,thick,inner sep=0pt,minimum width=1.5em]
\tikzstyle{ld} = [inner sep=1pt, font=\small]
\tikzstyle{unsite} = [circle, outer sep=0pt,inner sep=0.2pt,minimum width=0.75em]

\begin{document}
\title{Resolving competition of charge-density wave and superconducting phases using the MPS+MF algorithm}
\author{Gunnar Bollmark}
\affiliation{
	Department of Physics and Astronomy, Uppsala University, Box 516, S-751 20, Uppsala, Sweden
}
\author{Thomas K\"ohler}
\affiliation{
	Department of Physics and Astronomy, Uppsala University, Box 516, S-751 20, Uppsala, Sweden
}
\author{Adrian Kantian}
\affiliation{
	Department of Physics and Astronomy, Uppsala University, Box 516, S-751 20, Uppsala, Sweden
}
\affiliation{SUPA, Institute of Photonics and Quantum Sciences, Heriot-Watt University, Edinburgh EH14 4AS, United Kingdom}
\begin{abstract}
	Materials with strong electronic correlations may exhibit a superconducting phase when tuning some parameters, but they almost always also have multiple other phases, typically insulating ones, that are in close competition with superconductivity. It is highly challenging to resolve this competition with quantitative numerics for the group of quasi-two-dimensional (Q2D) materials such as the cuprates. This is the case even for the simplified minimal models of these materials, the doped 2D Hubbard model with repulsive interactions, where clusters of sufficient size to determine the phase in the thermodynamic limit can be hard-to-impossible to treat in practice. The present work shows how quasi-one-dimensional (Q1D) systems, 2D and 3D arrays of weakly coupled 1D correlated electrons, are much more amenable to resolve the competition between superconducting and insulating orders on an equal footing using many-body numerics. Using the recently established matrix product state plus mean field (MPS+MF) approach for fermions~\cite{Bollmark2022}, we demonstrate that large systems are readily reachable in these systems, and thus the thermodynamic regime by extrapolation. Focussing on basic model systems, 3D arrays of negative-$U$ Hubbard chains with additional nearest-neighbour interaction $V$, we show that despite the MF component of the MPS+MF technique we can reproduce the expected coexistence of superconductivity and charge-density wave at $V=0$ for density $n=1$. We then show how we can tune away from coexistence by both tuning $V$ and doping the system. This work thus paves the way to deploy two-channel MPS+MF theory on some highly demanding high-$T_c$ superconducting systems, such as 3D arrays of repulsive-$U$ doped Hubbard ladders, where we have recently characterized the properties of such arrays in single-channel MPS+MF calculations~\cite{Bollmark2022}. The present approach could thus conclusively show that this superconducting order would actually be obtained, by explicitly comparing superconductivity against it's insulating competitors (magnetic orders already being eliminated by design in these systems).
\end{abstract}
\maketitle

\section{\label{Sec::Introduction}Introduction}
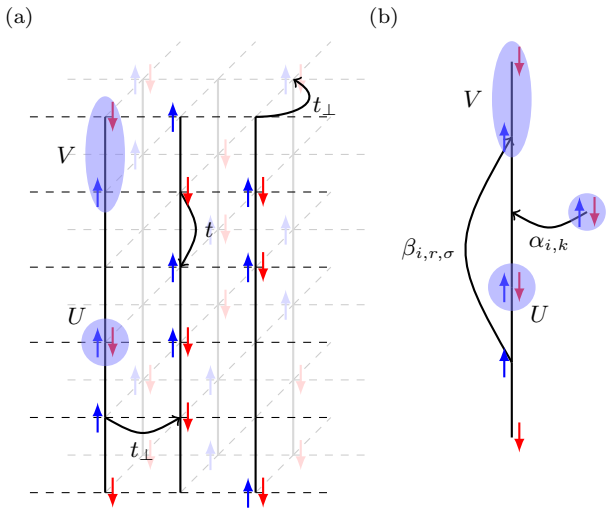
\begin{figure}
	\subfloat[\label{Fig::Full_model_schem}]{
	\tikzsetnextfilename{SchematicArrayFigure}
	\tikzset{external/export next=true}
	\centering
	\begin{tikzpicture}
	\foreach \y in {0,1,2,...,5}
	\foreach \x in {0,1,2}
	\foreach \o in {0,0.5}
	\draw[dashed,color=black!20] (\x+\o,\y+\o) -- (\x+0.5+\o,\y+0.5+\o);
	\foreach \y in {0,1,2,...,5}
	\foreach \x in {-1,0,1,2}
	\draw[dashed,color=black!20] (\x+0.5,\y+0.5) -- (\x+1+0.5,\y+0.5);
	\foreach \x in {0,1,2}
	\foreach \y in {0,1,2,...,4}
	\draw[color=black!15,thick] (0.5+\x,0.5+\y) -- (0.5+\x,0.5+\y+1);
	\foreach \x in {0,1,2}{
		\edef\shift{2*\x}
		\foreach \yi[evaluate=\yi as \y using {Mod(\yi+\shift,6)}] in {1,4,5}{
			\draw[-latex, thick, color=blue!15] (\x+0.4, \y+0.3) -- (\x+0.4,\y+0.7) node (b\x_\y) {};}
		\edef\shift{3*\x}
		\foreach \yi[evaluate=\yi as \y using {Mod(\yi+\shift,6)}] in {0,1,5}{
			\draw[-latex, thick, color=red!15] (\x+0.6, \y+0.7) -- (\x+0.6,\y+0.3) node (b\x_\y) {};
	}}
	\draw[->,thick] (2,5) to[out=0,in=-30,looseness=2, edge node={node[right,pos=0.5] {$t_\perp$}}](2.5,5.5);
	\draw[->,thick] (1,4) to[out=-60,in=60,looseness=1.4, edge node={node[right,pos=0.5] {$t$}}](1,3);
	\draw[->,thick] (0,1) to[out=-30,in=-150,looseness=1.4, edge node={node[below,pos=0.5] {$t_\perp$}}](1,1);
	\foreach \y in {0,1,2,...,5}
	\foreach \x in {-1,0,1,2}
	\draw[dashed] (\x,\y) -- (\x+1,\y);
	\foreach \x in {0,1,2}
	\foreach \y in {0,1,2,...,4}
	\draw[color=black!100,thick] (\x,\y) -- (\x,\y+1);
	\foreach \x in {0,1,2}{
		\edef\shift{1*\x}
		\foreach \yi[evaluate=\yi as \y using {Mod(\yi+\shift,6)}] in {1,2,4}{
	\draw[-latex, thick, color=blue] (\x-0.1, \y-0.2) -- (\x-0.1,\y+0.2) node (f\x_\y) {};}
		\edef\shift{1*\x+\x}
		\foreach \yi[evaluate=\yi as \y using {Mod(\yi+\shift,6)}] in {0,2,5}{
	\draw[-latex, thick, color=red] (\x+0.1, \y+0.2) -- (\x+0.1,\y-0.2) node (f\x_\y) {};
}}
	\node[circle,fill=blue,color=blue!50, opacity=0.5,minimum size=18] (vertex) at (0,2) {};
	\node[above left=-0.1 and -0.1 of vertex] {$U$};
	\node[ellipse,minimum height=44, minimum width=15, color=blue!50, opacity=0.5, fill] (nnvertex) at (0,4.5) {};
	\node[left=0 of nnvertex] {$V$};
	\end{tikzpicture}}
\subfloat[\label{Fig::Eff_1D_schem}]{
	\tikzsetnextfilename{Schematic1DFigure}
	\tikzset{external/export next=true}
	\begin{tikzpicture}
\draw[<-,thick] (0,3) to[out=-30,in=-150,looseness=1.4, edge node={node[below,pos=0.5] {$\alpha_{i,k}$}}](1,3);
\draw[->,thick] (0,1) to[out=120, in=240, looseness=1.4, edge node={node[left,pos=0.5] {$\beta_{i,r,\sigma}$}}] (0,4);
\foreach \x in {0}
\foreach \y in {0,1,2,...,4}
\draw[color=black!100,thick] (\x,\y) -- (\x,\y+1);
\foreach \x in {0}{
	\edef\shift{2*\x}
	\foreach \yi[evaluate=\yi as \y using {Mod(\yi+\shift,6)}] in {1,2,4}{
		\draw[-latex, thick, color=blue] (\x-0.1, \y-0.2) -- (\x-0.1,\y+0.2) node (f\x_\y) {};}
	\edef\shift{3*\x+\x}
	\foreach \yi[evaluate=\yi as \y using {Mod(\yi+\shift,6)}] in {0,2,5}{
		\draw[-latex, thick, color=red] (\x+0.1, \y+0.2) -- (\x+0.1,\y-0.2) node (f\x_\y) {};
}}
\draw[-latex, thick, color=red] (1+0.1, 3+0.2) -- (1+0.1,3-0.2) node (pd) {};
\draw[-latex, thick, color=blue] (1-0.1, 3-0.2) -- (1-0.1,3+0.2) node (pu) {};
\fill[color=blue!50,opacity=0.5] (1,3) circle (0.25) node (pair) {};
\node[circle,fill=blue,color=blue!50, opacity=0.5,minimum size=18] (vertex) at (0,2) {};
\node[below right=-0.1 and -0.1 of vertex] {$U$};
\node[ellipse,minimum height=44, minimum width=15, color=blue!50, opacity=0.5, fill] (nnvertex) at (0,4.5) {};
\node[left=0 of nnvertex] {$V$};
\end{tikzpicture}
}
	\caption{A schematic appearance of: (a) the full Hamiltonian given in \cref{full_ham} and (b) the effectively one-dimensional mean-field Hamiltonian of \cref{mf_ham}.}
	\label{Fig::Schematic}
\end{figure}
Understanding the macroscopic physics of strongly correlated electrons remains difficult for many systems. One of the main issues is the necessity of properly taking correlations between electrons into account, beyond effective single-particle descriptions. 
Problems where electronic correlations play a major role feature interesting and exotic physics ranging from \gls{USC} to spin-charge separation to \gls{Q1D} systems~\cite{BookAnderson1997,Stewart2017,Giamarchi2003}. 

A prominent example of the difficulty of accounting for correlations can be found in the \gls{2D} Hubbard models with repulsive interactions, originally proposed to describe the physics of $\mathrm{CuO}_2$ sheets in cuprate superconductors~\cite{Arovas2022}. This class of superconducting materials attain superconductivity at large critical temperatures when doped away from unit density. Yet, trying to match the complex known phase diagram of, e.g., the cuprates with that of a doped \gls{2D} Hubbard model at strong electron-electron repulsion remains very difficult~\cite{BookAnderson1997,Orenstein2000,Scalapino2012a,LeBlanc2015,Zheng2017,Bohrdt2019,Qin2020,Wietek2021,Bohrdt2021a}. This is due to the challenge of being able to treat a cluster of sufficient size to draw meaningful inferences about the thermodynamic limit of the system, which typically requires to resolve the competition of insulating and superconducting phases occurring for such systems~\cite{Arovas2022}. For techniques based on \gls{QMC}, this is due to an apparently incurable sign-problem, and for \gls{MPS}-based ones on the superlinear scaling of entanglement with subsystem surface~\cite{Stoudenmire2012}. Just for one particular doped \gls{2D} Hubbard model (pure square lattice with only on-site repulsion) it has taken a sustained multi-group, multi-method effort over many years to accrue evidence that it's ground state is more likely than not a stripe-order insulator instead of a $d$-wave superconductor~\cite{LeBlanc2015,Zheng2017,Qin2020,Wietek2021}. How these findings might change for any change in band structure, or for longer-range repulsion, or when coupling many \gls{2D} systems into a 3D array as required to represent an actual solid, is then another completely open problem.

\Gls{Q1D} many-body systems of fermions, which are \gls{2D} or 3D arrays of weakly coupled correlated \gls{1D} fermions, offer a much better outlook for efficiently resolving the competition between competing ordered phases in unconventionally and high-$T_c$ superconducting systems. Recently, an algorithm for solving such \gls{Q1D} systems using a combination of \gls{MPS} numerics and \gls{MF} theory~\cite{Bollmark2020,Bollmark2022,Marten2022} has been developed. The \gls{MPS}+\gls{MF} algorithm used in these works either studies cases where no competition of phases could occur or the winning phase was surmised from effective field theory arguments available due to the \gls{1D} nature of the sub-units from which the \gls{2D} or 3D array is constituted~\cite{Giamarchi2003}. However, these arguments are predicated entirely on the fastest diverging susceptibility of the \gls{1D} sub-units, which are then re-summed within a random phase or \gls{MF} approximation to yield susceptibilities for the entire array. The reasoning is thus incomplete -- in practice, different susceptibilities have different prefactors, which can also tip the balance towards a particular phase. Yet these prefactors are not known quantitatively from the effective field theory. 

In the present work we introduce and test a multi-channel-\gls{MF} version of the \gls{MPS}+\gls{MF} framework for correlated-fermion \gls{Q1D} systems that can resolve the competition between \gls{SC} and \gls{CDW} order, which are treated on an equal footing. This is done without pre-supposing any particular structure in the multiple ordering channels (other than the superconducting order being of singlet-type). This general approach allows the self-consistent flow to minimize the global (free) energy across multiple channels at once. This allows the expanded \gls{MPS}+\gls{MF} framework to also detect phase coexistence where this is physically possible.

In order to accomplish this a simple attractive fermion model composed of negative-$U$ Hubbard chains weakly coupled together is studied. Such a system admits solution via other methods allowing us to determine whether the constructed algorithm performs in accordance with expectation. Thus, the paper is structured as follows: In \cref{Sec::Model} the model is presented and microscopic parameters defined. Continuing, the \gls{MF} theory used to turn the \gls{Q1D} system into an effective \gls{1D} one is derived in \cref{Sec::MFtheory}. Additionally, the new channel of potential \gls{CDW} ordering is presented and its behavior discussed. In \cref{Sec::Results} the model is studied in both a strong and weak attractive regime. Coexistence is investigated by varying microscopic parameters and doping that affect the balance of orders. Finally, conclusions are given in \cref{Sec::Conclusion}.

$\;$
\section{\label{Sec::Model}Model}
In order to produce a \gls{MF} framework capable of resolving competition of \gls{CDW} and \gls{SC} in \gls{Q1D} systems we focus on a simple model amenable both to analytical methods and exact numerical methods:
\begin{equation}\label{full_ham}
	\hat H = \hat H_0 - t_\perp \hat H_\perp \;.
\end{equation}
The basic Hamiltonian is a set of disconnected \gls{1D} systems, described by $\hat H_0$, arranged equidistantly from each other. These 1D systems are indexed by a set of 2D vectors $\lbrace\textbf{R}_n\rbrace$. The disconnected Hamiltonian is given by (suppressing the $\textbf{R}_n$ on each operator)
\begin{widetext}
\begin{equation}\label{1d_hamiltonian}
	\hat H_0 = \sum_{\lbrace\textbf{R}_n\rbrace}\hat H_{1\mathrm{D}}(\textbf{R}_n) = \sum_{\lbrace\textbf{R}_n\rbrace}\left\lbrace-t\sum_{i\sigma}\left[\hat c^\dagger_{i+1,\sigma}\hat c^\nodagger_{i,\sigma} + \mathrm{h.c.}\right] - |U|\sum_in_{i,\uparrow}\hat n_{i,\downarrow} + V\sum_{i}\hat n_i \hat n_{i+1} - \mu\sum_i \hat n_i\right\rbrace \;,
\end{equation}
\end{widetext}
and the perpendicularly connecting Hamiltonian is given by
\begin{equation}
	\hat H_\perp =\frac{1}{2}\sum_{i}\sum_{\lbrace\textbf{R}_n\rbrace}\sum_{\langle\textbf{R}_n,\textbf{R}_m\rangle} \hat c^\dagger_{i,\sigma}(\textbf{R}_n)\hat c^\nodagger_{i,\sigma}(\textbf{R}_m) + \mathrm{h.c.} \;.
\end{equation}
Notably, tunneling only takes place between nearest-neighbor pairs $\langle\textbf{R}_n,\textbf{R}_m\rangle$. Dealing with a Hamiltonian of this form has been done using the \gls{MPS}+\gls{MF} algorithm~\cite{Bollmark2022,Bollmark2020}. For strongly correlated fermions previous work using \gls{MPS}+\gls{MF} has assumed \gls{SC} correlations dominate the ground state. However, studying systems with unit filling or adding the finite $V$ in \cref{1d_hamiltonian} this is no longer necessarily true. This requires a formulation of \gls{MPS}+\gls{MF} with an additional MF channel for describing insulating phases.

\section{Charge-density waves with Mean-Field theory\label{Sec::MFtheory}}
\begin{figure*}[t!]
	\tikzsetnextfilename{StrongAttractCompetition}
	\tikzset{external/export next=true}
	\begin{tikzpicture}
	\begin{groupplot}
	[
	group style = 
	{
		group size=3 by 2,
		vertical sep		=	1.5mm,
		horizontal sep		=	1.5mm,
		x descriptions at	=	edge bottom,
		y descriptions at	=	edge left,
	},
	ymin	= -0.1,
	ymax	= 2.1,
	width=0.4\textwidth-3.75pt,
	height=0.2\textheight,
	xticklabel style = {/pgf/number format/precision=3,
	/pgf/number format/fixed},
	xlabel = {$i$},
	]
	\nextgroupplot
	[
	title = {$V=-0.05t$},
	ylabel = {$\braket{\hat c_{i,\uparrow} \hat c_{i,\downarrow}}$},
	ymin	= -0.025,
	ymax	= 0.425,
	]
	\addplot
	[
	red,
	thin,
	only marks,
	mark=|,
	]
	table
	[
	y expr = \thisrowno{1},
	x expr = \thisrowno{0}
	]
	{Data/last_orp_Pr_1_Phr_1_tp_0.3_L_96_n_1_U_-10.0_V_-0.05_chi_200.dat};
	\addplot
	[
	blue,
	thin,
	only marks,
	mark=|,
	]
	table
	[
	y expr = \thisrowno{1},
	x expr = \thisrowno{0}
	]
	{Data/prev_orp_Pr_1_Phr_1_tp_0.3_L_96_n_1_U_-10.0_V_-0.05_chi_200.dat};
	\nextgroupplot
	[
	title = {$V=0$},
	ymin	= -0.025,
	ymax	= 0.425,
	]
	\addplot
	[
	red,
	thin,
	only marks,
	mark=|,
	]
	table
	[
	y expr = \thisrowno{1},
	x expr = \thisrowno{0}
	]
	{Data/last_orp_Pr_1_Phr_1_tp_0.3_L_96_n_1_U_-10.0_V_0.0_chi_200.dat};
	\addplot
	[
	blue,
	thin,
	only marks,
	mark=|,
	]
	table
	[
	y expr = \thisrowno{1},
	x expr = \thisrowno{0}
	]
	{Data/prev_orp_Pr_1_Phr_1_tp_0.3_L_96_n_1_U_-10.0_V_0.0_chi_200.dat};
	\nextgroupplot
	[
	title = {$V=0.05t$},
	legend pos = north east,
	ymin	= -0.025,
	ymax	= 0.425,
	]
	\addplot
	[
	red,
	thin,
	only marks,
	mark=|,
	]
	table
	[
	y expr = \thisrowno{1},
	x expr = \thisrowno{0}
	]
	{Data/last_orp_Pr_1_Phr_1_tp_0.3_L_96_n_1_U_-10.0_V_0.05_chi_200.dat};
	\addlegendentry{Period 1}
	\addplot
	[
	blue,
	thin,
	only marks,
	mark=|,
	]
	table
	[
	y expr = \thisrowno{1},
	x expr = \thisrowno{0}
	]
	{Data/prev_orp_Pr_1_Phr_1_tp_0.3_L_96_n_1_U_-10.0_V_0.05_chi_200.dat};
	\addlegendentry{Period 2}
	\nextgroupplot
	[
	ylabel = {$\braket{\hat n_i}$},
	]
	\addplot
	[
	red,
	only marks,
	mark=|,
	]
	table
	[
	y expr = \thisrowno{1},
	x expr = \thisrowno{0}
	]
	{Data/last_dens_Pr_1_Phr_1_tp_0.3_L_96_n_1_U_-10.0_V_-0.05_chi_200.dat};
	\addplot
	[
	blue,
	only marks,
	mark=|,
	]
	table
	[
	y expr = \thisrowno{1},
	x expr = \thisrowno{0}
	]
	{Data/prev_dens_Pr_1_Phr_1_tp_0.3_L_96_n_1_U_-10.0_V_-0.05_chi_200.dat};
	\nextgroupplot
	[
	]
	\addplot
	[
	red,
	thin,
	only marks,
	mark=|,
	]
	table
	[
	y expr = \thisrowno{1},
	x expr = \thisrowno{0}
	]
	{Data/last_dens_Pr_1_Phr_1_tp_0.3_L_96_n_1_U_-10.0_V_0.0_chi_200.dat};
	\addplot
	[
	blue,
	thin,
	only marks,
	mark=|,
	]
	table
	[
	y expr = \thisrowno{1},
	x expr = \thisrowno{0}
	]
	{Data/prev_dens_Pr_1_Phr_1_tp_0.3_L_96_n_1_U_-10.0_V_0.0_chi_200.dat};
	\nextgroupplot
	[
	]
	\addplot
	[
	red,
	only marks,
	mark=|,
	]
	table
	[
	y expr = \thisrowno{1},
	x expr = \thisrowno{0}
	]
	{Data/last_dens_Pr_1_Phr_1_tp_0.3_L_96_n_1_U_-10.0_V_0.05_chi_200.dat};
	\addplot
	[
	blue,
	thin,
	only marks,
	mark=|,
	]
	table
	[
	y expr = \thisrowno{1},
	x expr = \thisrowno{0}
	]
	{Data/prev_dens_Pr_1_Phr_1_tp_0.3_L_96_n_1_U_-10.0_V_0.05_chi_200.dat};
	\end{groupplot}
	\end{tikzpicture}
	\caption{Figure showing the competition between \gls{SC} and \gls{CDW} phases at $t_\perp=0.3t$ and $U=-10t$ for a system of length $L=96$. The upper row shows a plot of local pairing amplitude over system size while the bottom row shows local density. In each figure the two final self-consistency iterations are plotted: When the plot is blue it lies on top of the red one. If both iterations are visible then they are dissimilar indicating a \gls{CDW} phase.}
	\label{Fig::Competition}
\end{figure*}
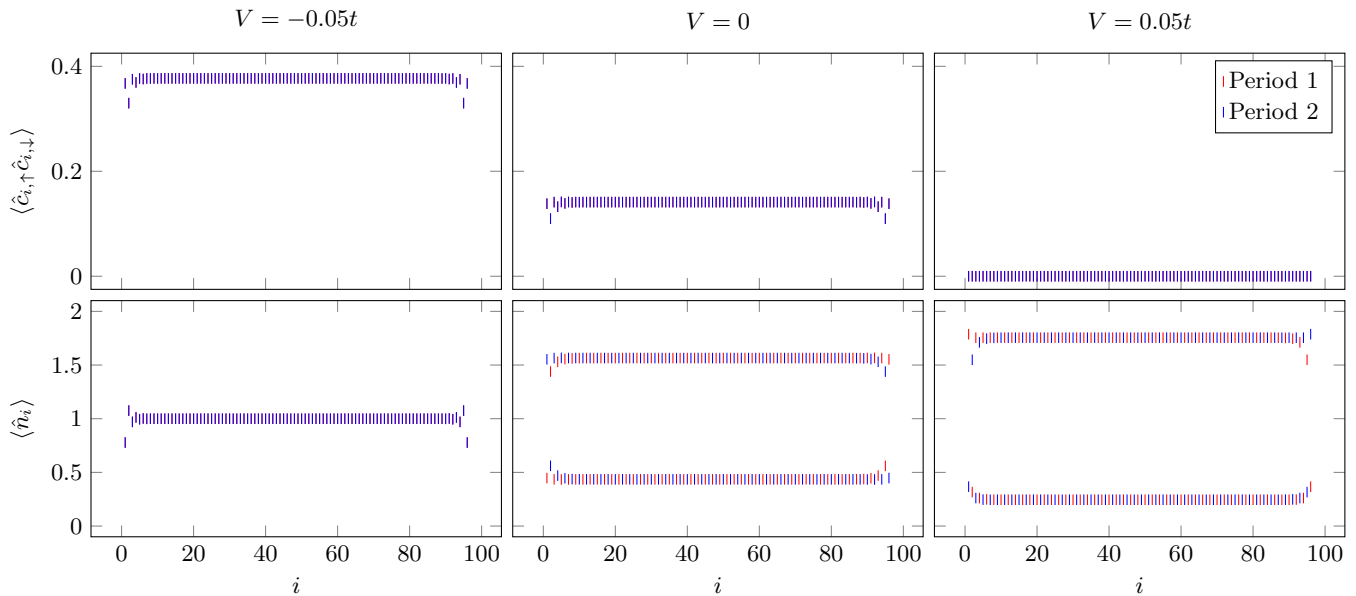
In deriving the self-consistent equations for \gls{MPS}+\gls{MF} a perturbation theory is first applied to turn tunneling into effective interaction. Subsequently, a MF treatment is applied to this interaction: The usual Hartree-Fock and Bogoliubov contributions are taken into account producing a self-consistent effective \gls{1D} Hamiltonian~\cite{Bollmark2022}
\begin{align}\label{mf_ham}
\nonumber \hat H_{\mathrm{MF}} = & \hat H_{1\mathrm{D}}(\textbf{R}_n) - \sum_{i,k}\alpha_{i,k}\left( \hat c^{\nodagger}_{i,\uparrow} \hat c^{\nodagger}_{k,\downarrow} + \hat c^\dagger_{k,\downarrow} \hat c^\dagger_{i,\uparrow} \right) \\
&+ \sum_{i,\sigma}\sum_{r=1}^{L-i} \beta_{i,r,\sigma}\left(\hat c^\dagger_{i+r,\sigma} \hat c^{\nodagger}_{i,\sigma} + \hat c^\dagger_{i,\sigma} \hat c^{\nodagger}_{i+r,\sigma}\right) \\ &+ \sum_{i,\sigma}\mu_{i,\sigma}\hat n_{i,\sigma} \;,
\end{align}
where
\begin{align}
\alpha_{i,k} = \frac{2 z_\mathrm{c} t_\perp^2}{\Delta E_\mathrm{p}} \braket{\hat c^{\nodagger}_{i,\uparrow} \hat c^{\nodagger}_{k,\downarrow}}\;,\label{alphas}\\
\beta_{i,r,\sigma} = \frac{2 z_\mathrm{c} t_\perp^2}{\Delta E_\mathrm{p}}
\label{betas}
\braket{\hat c^\dagger_{i+r,\sigma} \hat c^{\nodagger}_{i,\sigma}}\; \\
\mu_{i,\sigma} = \frac{z_ct_\perp^2}{\Delta E_p}(2\braket{\hat n_{i,\sigma}}-1)\; ,\label{mu_sc}
\end{align}
where all amplitudes are assumed to be real. This requires the definition of pairing energy
\begin{align}
	\Delta E_\mathrm{p} = 2E\left(N+1,S=\frac{1}{2}\right)-E\left(N+2,S=0\right)\nonumber\\-E(N,S=0) \;,
\end{align}
with $E(N,S)$ the spectrum of $\hat H_{1\mathrm{D}}$ indexed by particle number and spin. The coordination number $z_\mathrm{c}$ count the neighboring systems of a single \gls{1D} sub-system.

Notably, the \gls{MF} terms Eqs.~\eqref{alphas}-\eqref{mu_sc} are easily interpreted: Pairing lowers system energy whereas particle-hole terms carry a cost instead. Thus, pairing with any separation of the constituents is generally favored whereas dispersion is disfavored. The final term is unique in that it provides a repulsive interaction, capable of driving charge order. At the same time, density is constant for a system which orders in the pairing operator and $\mu_{i,\sigma}$ would simply modify the chemical potential $\mu$. In this work we expand the \gls{MF} theory to include a fully site-dependent $\mu_{i,\sigma}$-term, which can drive charge-ordering.

Solving such simplified systems is the subject of \gls{MPS}+\gls{MF} and is done by iteration of the \gls{MF} amplitudes until they achieve self-consistency. While, in general the initial configuration of these amplitudes should not affect the final solution, in practice there are several choices that are more advantageous than others as expounded on in \cref{App::Initial}.

At the basis of the MF Hamiltonian in Eq.~\eqref{mf_ham} lies perturbation theory which produces an effective inter-chain interaction. It is notable that one form of interaction produced is a repulsive one which will tend to disfavor neighboring particles of like spin in the transverse direction. Thus, it would lead to a pinning behavior in these directions yielding a checkerboard appearance. Hence, we call such a phase \gls{CDW} inspired by the \gls{1D} physics of $\hat H_{1\mathrm{D}}$~\cite{Giamarchi2003}. In order to solve this expanded set of amplitudes we must answer how such a term behaves under the \gls{MF} theory used in \gls{MPS}+\gls{MF}.
\subsection{Two-period in mean-fields}
Performing \gls{MF} theory on the inter-chain interaction changes the meaning of the term. While, in general, such interaction would compare densities at different chains, the \gls{MF} treatment assumes that all \gls{1D} sub-systems are the same, turning the transverse interaction into a density comparison in the same chain. This is undesireable since it gives rise to a local interaction between same-spin particles which would be prohibited by the Pauli principle. However, numerically solving for self-consistency requires iterative solution. Hence, the energy contribution to minimize at each iteration is really given by
\begin{equation}
	V_{\perp} \propto (2\braket{\hat n_{i,\sigma}}_{j-1}-1)\braket{\hat n_{i,\sigma}}_{j}\;,
\end{equation}
where $\braket{}_j$ indicates measurement at iteration $j$ of the self-consistent calculation. Notably, since the interaction is repulsive the system will strive to minimize $\braket{\hat n_{i,\sigma}}_{j-1}\braket{\hat n_{i,\sigma}}_j$. In other words, a state ordering in this operator will in iteration $j$ avoid high density in the regions of iteration $j-1$ which already had high density there. Hence a checkerboard-like pattern is achieved.

The cost of this is that the self-consistent iteration will at best converge to two different states, endlessly oscillating between the two so as to keep $\braket{\hat n_{i,\sigma}}_{j-1}\braket{\hat n_{i,\sigma}}_j$ small. When this occurs we name it a two-periodic behavior and surmise that it indicates that ordering in the \gls{CDW} channel of the system is preferred. One concern of this interpretation is whether the Hamiltonian maintains tunneling equally strong to and from sites of large and small density. Since the MF amplitudes in Eq.~\eqref{betas} depend on this measurement it may be a concern whether hermiticity is maintained. We reason that this is not an issue case due to high-amplitude transitions always end up in low-amplitude states and vice versa (see \cref{App::Hermiticity}).
\section{\label{Sec::Results}Results}

In this work we are primarily interested in whether and how an MF theory based on the two-periodic MF amplitude in \cref{Sec::MFtheory} can resolve physics expected from analytical methods in the same system. As such, we first study strongly interacting fermions at $U=-10t$ in a unit filled system
\begin{equation}
	n = \frac{1}{L}\sum_{i=1}^L(\braket{\hat n_{i,\uparrow}}+\braket{\hat n_{i,\downarrow}}) = 1\;.
\end{equation}
The pairing gap in such a system we compute to approximately $\Delta E_p\approx6.5t$ yielding tightly bound pairs. This allows us to cut off the number of mean-field amplitudes (this is explained in more detail in~\cite{Bollmark2022})
\begin{align}
	\alpha_{i,k} =
	\begin{cases}
	\alpha_{i,k} & \text{if } |i-k| \leq R \\
	0 & \text{otherwise}
	\end{cases},
	\\
	\beta_{i,r,\sigma} =
	\begin{cases}
	\beta_{i,r,\sigma} & \text{if } |i-k| \leq R \\
	0 & \text{otherwise}
	\end{cases}\;,
\end{align}
where $R$ is the cut-off point. With very tight binding such as for $U=-10t$ we choose this value at $R=1$. The great advantage of this system is that it is amenable to practically exact treatment which can provide useful standards of comparison to the developed method.

Once the method is established we attempt to develop a methodology in a lightly doped system at weaker interaction of $U=-2t$. In this case the pairing energy is somewhat affected by the other microscopic parameters and we find that it evaluates to $\Delta E_p \in [0.1t,0.2t]$. Previous work on these systems indicate that a range of $R=5$ should yield acceptable precision and as such we limit simulation to $R=5$ when $U=-2t$. Interestingly, such a weakly attractive system lies close in similarity to previously studied lightly doped Hubbard ladders~\cite{Bollmark2022}. Therefore we also study the effects of doping in the weakly attractive case to see its effects on competition.

Notably, since the algorithm solving the effective \gls{1D} problem at each iteration is a \gls{MPS}-based implementation of \gls{DMRG} it is subject to the precision limitations of such an algorithm~\cite{White1992,White1993,Schollwock2005,Schollwock2011}. The two parameters which affect precision are system size and bond dimension which are considered in \cref{App::Precision}. It is found that the precision is not sufficiently affected by these parameters to change the qualitative nature of the results. Thus, the sizes and bond dimensions which are found in \cref{App::Precision} are presented in this section.

\subsection{Unit filling competition}
At unit density the \gls{CDW} and \gls{SC} phases can be shown to coexist analytically by virtue of a particle-hole transformation~\cite{Ho2009}. Specifically, there is symmetry between \gls{CDW} and \gls{SC} phases in such conditions meaning that if one occurs then the other is also present by symmetry. Any method aiming to resolve competition between these phases should be able to find such coexistence at unit density. In order to achieve this we focus on strongly attracting fermions. For such a case there is particularly good control as fermionic particles pair up more often than not leading to a bosonic description. 

In order to resolve the competition we study the \gls{MF} amplitudes corresponding to \gls{CDW} and \gls{SC} order respectively. For \gls{CDW} order the local density should enter a two-periodic behavior when the system is minimizing energy using this channel. Conversely, a superconductor can be studied with the \gls{BCS} order parameter $\braket{\hat c_{i,\uparrow} \hat c_{i,\downarrow}}$. Whether the local density enters a two-periodic behavior or the \gls{BCS} order parameter attains finite value will determine which order the ground state obtains.

Knowing what to look for, the nearest neighbor interaction $V$ is varied in order to bias the system towards any given order: repulsion, $V>0$, for \gls{CDW} and attraction, $V<0$ for \gls{SC}. As can be seen in \cref{Fig::Competition} the system's ground state changes notably over a small range of nearest neighbor interaction. The appearance can be roughly summarized as follows:
\begin{enumerate}
	\item Weak attraction: The systems local density approaches a constant throughout the system and likewise the local pairing.
	\item No interaction: Local density enters a two-periodic behavior while simultaneously local pairing remains in a single period and constant throughout the system.
	\item Weak repulsion: Local pairing is now zero and local density has a strong two-period.
\end{enumerate}

These results indicate that \gls{MPS}+\gls{MF} can be used to meaningfully study competition of \gls{CDW} and \gls{SC} phases in the model under study. Notably, the strongly interacting regime is amenable to other methods: particularly due to its likeness to a bosonic model where each doublon represent a single hard-core boson.

While it is important to keep the structure of the solution provided in \cref{Fig::Competition} in mind it contains a lot more information than is necessary for determination of phases. The superconducting order parameter is approximately constant throughout the system inviting the definition of a system-wide order parameter
\begin{equation}\label{sc_order_par}
	\braket{\hat c \hat c}=\frac{1}{i_l-i_f+1}\sum_{i=i_f}^{i_l}\braket{\hat c_{i,\uparrow} \hat c_{i,\downarrow}}\;.
\end{equation}
The inclusion of limits to the average is due to the \gls{OBC} nature of \gls{DMRG}: the edges act as impurities which disturb the bulk solution in which we are truly interested. In this manner we can eliminate edge effects by averaging over the center of the system.

\begin{figure}
	\tikzsetnextfilename{OrpHeatMap}
	\tikzset{external/export next=true}
	\centering
	\begin{tikzpicture}
	\foreach \V [count=\n] in {-0.05,0.0,0.05}
	\draw[thick] (0.9,\n+0.5) to[edge node={node[left,pos=0] {\V}}](1.1,\n+0.5);
	\foreach \tp [count=\n]in {0.2,0.3,0.4,0.5}{
		\draw[thick] (\n+0.5,1.1) to[edge node={node[below,pos=1] {\tp}}](\n+0.5,0.9);
		\foreach \V [count=\m] in {-0.05,0.0,0.05}{
			\pgfplotstableread[header=false]{Data/sc_orp_Pr_1_Phr_1_tp_\tp_L_160_n_1_U_-10.0_V_\V_chi_200.dat}{\fopsc}
			\pgfplotstablegetelem{0}{[index]0}\of\fopsc
			\pgfmathsetmacro\sccolval{100*\pgfplotsretval/0.46}
			\pgfplotstableread[header=false]{Data/cdw_orp_Pr_1_Phr_1_tp_\tp_L_160_n_1_U_-10.0_V_\V_chi_200.dat}{\fopcdw}
			\pgfplotstablegetelem{0}{[index]0}\of\fopcdw
			\pgfmathsetmacro\cdwcolval{100*\pgfplotsretval/2}
			\fill[color=red!\sccolval!white] (\n,\m) -- (\n+1,\m) -- (\n,\m+1);
			\fill[color=blue!\cdwcolval!white] (\n,\m+1) -- (\n+1,\m+1) -- (\n+1,\m);}}
	\draw[->,thick] (1,1) to[edge node={node[left=20pt,pos=0.4287] {$V/t$}}](1,4.5);
	\draw[->,thick] (1,1) to[edge node={node[below=20pt,pos=0.4545] {$t_\perp/t$}}](5.5,1);
	\draw[fill,shading=axis,top color=red,bottom color=white] (6.3,0.5) to (6.5,0.5) to (6.5,4.5) to (6.3,4.5) to[edge node={node[left=6pt,pos=0.5,rotate=90,anchor=center] {SC}}] (6.3,0.5);
	\draw[fill,shading=axis,top color=blue,bottom color=white] (6.8,0.5) to (7,0.5) to[edge node={node[right=6pt,pos=0.5,rotate=90,anchor=center] {CDW}}] (7,4.5) to (6.8,4.5) to (6.8,0.5);
	\draw (6.2,0.5) to[edge node={node[left,pos=0] {0}}](6.3,0.5);
	\draw (6.2,4.5) to[edge node={node[left,pos=0] {0.46}}](6.3,4.5);
	\draw (7,0.5) to[edge node={node[right,pos=1] {0}}](7.1,0.5);
	\draw (7,4.5) to[edge node={node[right,pos=1] {2}}](7.1,4.5);
	\end{tikzpicture}
	\caption{Order parameter of \gls{SC} and \gls{CDW} order from \cref{sc_order_par,cdw_order_par}, respectively. The simulation is performed for $U=-10t$, $n=1$, $L=160$ and varying $t_\perp$ and $V$. The \gls{SC} order uses red color and \gls{CDW} order blue.}
	\label{Fig::Order_par_heatmap}
\end{figure}
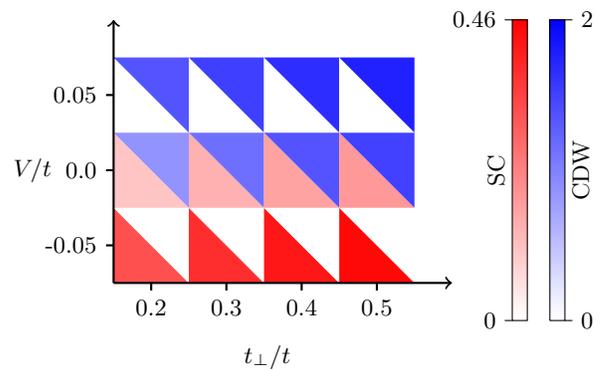

Typically, a \gls{CDW} phase would be identified by the long-range behavior of density-density correlations. However, when the term which drives \gls{CDW} orders, the self-consistent solution provides two different ground states and self-consistent Hamiltonians. A linear combination of these ground states may solve the issue but requires in general that the states do not overlap: Something we cannot guarantee. Instead, it is clear that the two-periodic behavior can be diagnosed by a difference of local density at two different self-consistent iterations:
\begin{equation}\label{cdw_order_par}
	\delta n = \frac{1}{i_l-i_f+1}\sum_{i=i_f}^{i_l}|\braket{\hat n_i}_{j}-\braket{\hat n_i}_{j-1}| \;.
\end{equation}
Using these definitions a mapping of the order parameters at several different $V$ and $t_\perp$ can be easily displayed as shown in \cref{Fig::Order_par_heatmap}.

While being able to qualitatively replicate known results is useful as verification for the method, we can use the \gls{MPS}+\gls{MF} routine to move on to models of weakly interacting fermions. Moving outside the composite bosonic regime we target a weak attraction of $U=-2t$. Such a system has comparable pairing strength to weakly doped repulsive Hubbard ladders, making for an interesting study on the effects of competition in such systems.

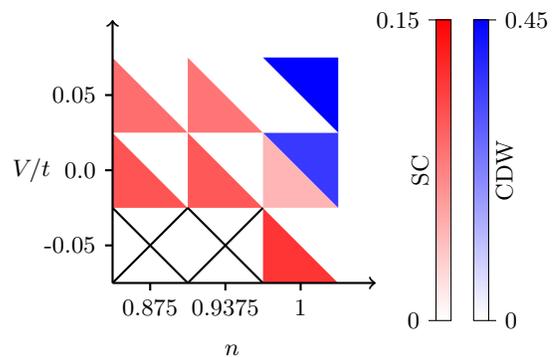
\begin{figure}
	\tikzsetnextfilename{OrpHeatMapWeakAttract}
	\tikzset{external/export next=true}
	\centering
	\begin{tikzpicture}
	\foreach \V [count=\n] in {-0.05,0.0,0.05}
	\draw[thick] (0.9,\n+0.5) to[edge node={node[left,pos=0] {\V}}](1,\n+0.5);
	\foreach \d [count=\n]in {0.875,0.9375,1}{
		\draw[thick] (\n+0.5,1) to[edge node={node[below,pos=1] {\d}}](\n+0.5,0.9);
		\foreach \V [count=\m] in {0.0,0.05}{
			\pgfplotstableread[header=false]{Data/sc_orp_Pr_5_Phr_5_tp_0.05_L_192_n_\d_U_-2.0_V_\V_chi_200.dat}{\fopsc}
			\pgfplotstablegetelem{0}{[index]0}\of\fopsc
			\pgfmathsetmacro\sccolval{100*\pgfplotsretval/0.15}
			\pgfplotstableread[header=false]{Data/cdw_orp_Pr_5_Phr_5_tp_0.05_L_192_n_\d_U_-2.0_V_\V_chi_200.dat}{\fopcdw}
			\pgfplotstablegetelem{0}{[index]0}\of\fopcdw
			\pgfmathsetmacro\cdwcolval{100*\pgfplotsretval/0.45}
			\fill[color=red!\sccolval!white] (\n,\m+1) -- (\n+1,\m+1) -- (\n,\m+2);
			\fill[color=blue!\cdwcolval!white] (\n,\m+2) -- (\n+1,\m+2) -- (\n+1,\m+1);}}
	\pgfplotstableread[header=false]{Data/sc_orp_Pr_5_Phr_5_tp_0.05_L_192_n_1_U_-2.0_V_-0.05_chi_200.dat}{\fopsc}
	\pgfplotstablegetelem{0}{[index]0}\of\fopsc
	\pgfmathsetmacro\sccolval{100*\pgfplotsretval/0.15}
	\pgfplotstableread[header=false]{Data/cdw_orp_Pr_5_Phr_5_tp_0.05_L_192_n_1_U_-2.0_V_-0.05_chi_200.dat}{\fopcdw}
	\pgfplotstablegetelem{0}{[index]0}\of\fopcdw
	\pgfmathsetmacro\cdwcolval{100*\pgfplotsretval/0.45}
	\fill[color=red!\sccolval!white] (3,1) -- (4,1) -- (3,2);
	\fill[color=blue!\cdwcolval!white] (3,2) -- (4,2) -- (4,1);
	\draw[thick] (1,1) to (2,2);
	\draw[thick] (1,2) to (2,1);
	\draw[thick] (2,1) to (3,2);
	\draw[thick] (2,2) to (3,1);
	\draw[->,thick] (1,1) to[edge node={node[left=20pt,pos=0.4287] {$V/t$}}](1,4.5);
	\draw[->,thick] (1,1) to[edge node={node[below=20pt,pos=0.4545] {$n$}}](4.5,1);
	\draw[fill,shading=axis,top color=red,bottom color=white] (5.3,0.5) to (5.5,0.5) to (5.5,4.5) to (5.3,4.5) to[edge node={node[left=6pt,pos=0.5,rotate=90,anchor=center] {SC}}] (5.3,0.5);
	\draw[fill,shading=axis,top color=blue,bottom color=white] (5.8,0.5) to (6,0.5) to[edge node={node[right=6pt,pos=0.5,rotate=90,anchor=center] {CDW}}] (6,4.5) to (5.8,4.5) to (5.8,0.5);
	\draw (5.2,0.5) to[edge node={node[left,pos=0] {0}}](5.3,0.5);
	\draw (5.2,4.5) to[edge node={node[left,pos=0] {0.15}}](5.3,4.5);
	\draw (6,0.5) to[edge node={node[right,pos=1] {0}}](6.1,0.5);
	\draw (6,4.5) to[edge node={node[right,pos=1] {0.45}}](6.1,4.5);
	\end{tikzpicture}
	\caption{Order parameter of \gls{SC} and \gls{CDW} order from \cref{sc_order_par,cdw_order_par}, respectively. The simulation is performed for for $U=-2t$, $t_\perp=0.05t$, $L=192$ and varying $n$ and $V$. The \gls{SC} order uses red color and \gls{CDW} order blue.}
	\label{Fig::weak_order_par_heatmap}
\end{figure}

Notably, we find the qualitative picture remains: \Cref{Fig::weak_order_par_heatmap} shows the competition when moving away from $V=0$. The coexistence is found once again for $V=0$ in the same figure.

\subsection{Hole doping}
In previous studies the \gls{MPS}+\gls{MF} method has been applied to systems which exhibit \gls{USC} as a consequence of repulsively mediated electron pairing~\cite{Bollmark2022}. Specifically, lightly doped repulsive Hubbard ladders weakly coupled in a \gls{Q1D} array are known to feature a quasi-long range order instability to \gls{SC} but simultaneously to that of \gls{CDW}. This competition is currently not possible to conclusively resolve with analytical methods and a more quantitative method would be the safest alternative.

In \cref{Fig::weak_order_par_heatmap} we show an example of a weakly attracting system doped away from unit filling. Slightly doping the system to a filling of $n=15/16$ sees the local pairing and superconductivity remain in the system while order parameter even increases. Interestingly, the local density difference is clearly finite at $n=1$ with a two-periodic behavior that disappears with doping. Furthermore, applying a repulsive potential to the system as when $n=1$, thereby increasing the cost of pair condensation no longer destroys the superconductor.

\section{Conclusions\label{Sec::Conclusion}}
In this work we have developed a modified version of the existing \gls{MPS}+\gls{MF} framework. Lifting assumptions of dominant ordering the competition between \gls{SC} and \gls{CDW} phases in fermionic systems with heuristic pairing is studied. This method carries the benefit of resolving competition in a \gls{MF} sense without the use of Renormalization Group theory.

From the developed method it is apparent that the expectation of coexistence of \gls{CDW} and \gls{SC} phases at unit filling can be resolved. Including nearest neighbor interactions allows the biasing of the system towards either a \gls{SC} or \gls{CDW} phase. With weak bias to either attraction or repulsion we find the coexistence broken in the expected direction. Thus, we conclude that the method is able to resolve competition between these two phases at the very least on the \gls{MF} level.

Utilizing the newly developed method we apply it to a more complicated case: Weakly interacting fermions doped away from unit filling. In this case, doping lifts the coexistence and favors a \gls{SC} phase: Something that is also concluded by quasi-ordering arguments in bosonization calculations.

For future work, this method can be utilized in resolving competition between phases in more complex systems such as, e.g., arrays of doped repulsive Hubbard ladders: Conclusively answering the question of which phase such a system orders in.

\begin{acknowledgments}
	This work was supported by an ERC Starting Grant from the European Union's Horizon 2020 research and innovation programme under grant agreement No. 758935; and the UK's Engineering and Physical Sciences Research Council [grant number EP/W022982/1]. The computations were enabled by resources provided by the Swedish National Infrastructure for Computing (SNIC), as well as by compute time awarded by the UK’s Engineering and Physical Sciences Research Council on the ARCHER2, Peta4-Skylake and Cirrus compute clusters through the ``Access to HPC" and ``Scottish Academic Access" calls.
	
\end{acknowledgments}
\newpage
\appendix
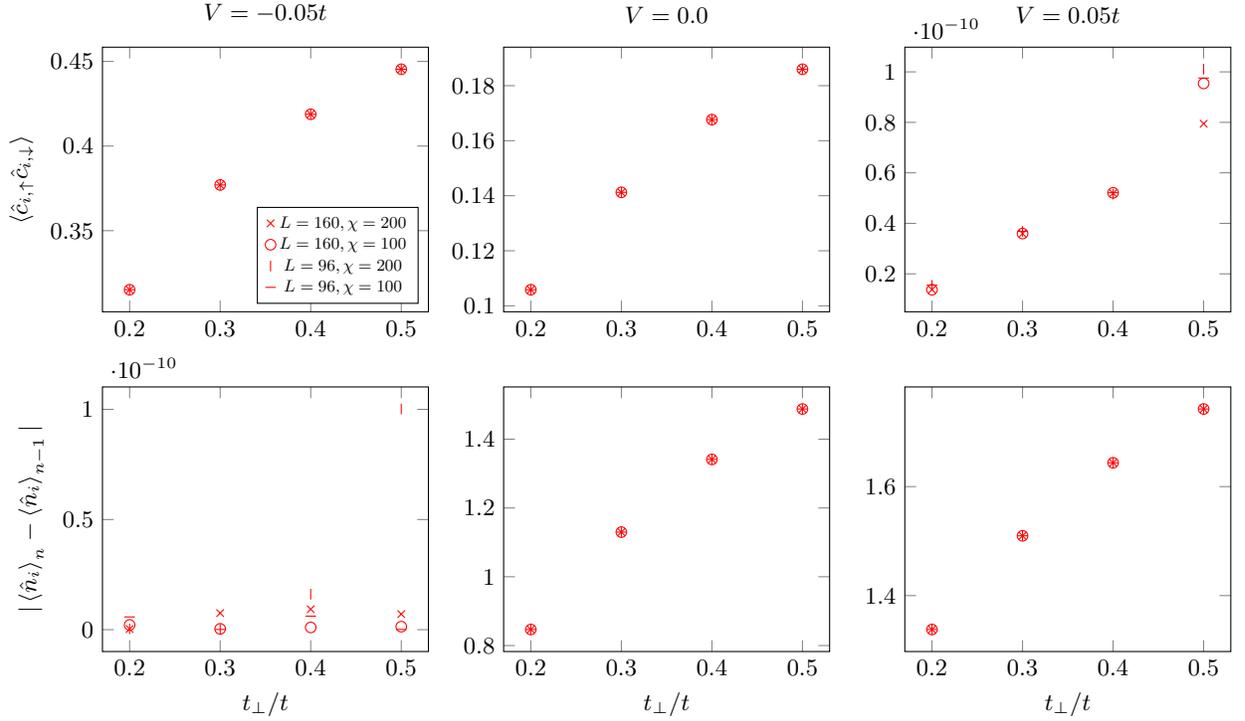
\begin{figure*}[t!]
	\tikzsetnextfilename{PrecisionQualityCheck10U}
	\tikzset{external/export next=true}
	\centering
	\begin{tikzpicture}
	\begin{groupplot}
	[group style = {group size=3 by 2}]
	\nextgroupplot
	[
	width=0.33\textwidth,
	title = {$V=-0.05t$},
	ylabel = {$\braket{\hat c_{i,\uparrow} \hat c_{i,\downarrow}}$},
	legend pos = south east,
	legend style={nodes={scale=0.7, transform shape}},
	xticklabel style = {/pgf/number format/precision=3,
		/pgf/number format/fixed},
	yticklabel style = {/pgf/number format/precision=3,
		/pgf/number format/fixed},
	scaled y ticks = false,
	]
	\addplot
	[
	red,
	thin,
	only marks,
	mark=x,
	]
	table
	[
	y expr = \thisrowno{1},
	x expr = \thisrowno{0},
	]
	{Data/sc_orp_over_tp_Pr_1_Phr_1_L_160_n_1_U_-10.0_V_-0.05_chi_200.dat};
	\addlegendentry{$L=160,\chi=200$}
	\addplot
	[
	red,
	thin,
	only marks,
	mark=o,
	]
	table
	[
	y expr = \thisrowno{1},
	x expr = \thisrowno{0},
	]
	{Data/sc_orp_over_tp_Pr_1_Phr_1_L_160_n_1_U_-10.0_V_-0.05_chi_100.dat};
	\addlegendentry{$L=160,\chi=100$}
	\addplot
	[
	red,
	thin,
	only marks,
	mark=|,
	]
	table
	[
	y expr = \thisrowno{1},
	x expr = \thisrowno{0},
	]
	{Data/sc_orp_over_tp_Pr_1_Phr_1_L_96_n_1_U_-10.0_V_-0.05_chi_200.dat};
	\addlegendentry{$L=96,\chi=200$}
	\addplot
	[
	red,
	thin,
	only marks,
	mark=-,
	]
	table
	[
	y expr = \thisrowno{1},
	x expr = \thisrowno{0},
	]
	{Data/sc_orp_over_tp_Pr_1_Phr_1_L_96_n_1_U_-10.0_V_-0.05_chi_100.dat};
	\addlegendentry{$L=96,\chi=100$}
	\nextgroupplot
	[
	width=0.33\textwidth,
	title = {$V=0.0$},
	legend pos = south east,
	legend style={nodes={scale=0.7, transform shape}},
	xticklabel style = {/pgf/number format/precision=3,
		/pgf/number format/fixed},
	yticklabel style = {/pgf/number format/precision=3,
		/pgf/number format/fixed},
	scaled y ticks = false,
	]
	\addplot
	[
	red,
	thin,
	only marks,
	mark=x,
	]
	table
	[
	y expr = \thisrowno{1},
	x expr = \thisrowno{0},
	]
	{Data/sc_orp_over_tp_Pr_1_Phr_1_L_160_n_1_U_-10.0_V_0.0_chi_200.dat};
	\addplot
	[
	red,
	thin,
	only marks,
	mark=o,
	]
	table
	[
	y expr = \thisrowno{1},
	x expr = \thisrowno{0},
	]
	{Data/sc_orp_over_tp_Pr_1_Phr_1_L_160_n_1_U_-10.0_V_0.0_chi_100.dat};
	\addplot
	[
	red,
	thin,
	only marks,
	mark=|,
	]
	table
	[
	y expr = \thisrowno{1},
	x expr = \thisrowno{0},
	]
	{Data/sc_orp_over_tp_Pr_1_Phr_1_L_96_n_1_U_-10.0_V_0.0_chi_200.dat};
	\addplot
	[
	red,
	thin,
	only marks,
	mark=-,
	]
	table
	[
	y expr = \thisrowno{1},
	x expr = \thisrowno{0},
	]
	{Data/sc_orp_over_tp_Pr_1_Phr_1_L_96_n_1_U_-10.0_V_0.0_chi_100.dat};
	\nextgroupplot
	[
	width=0.33\textwidth,
	title = {$V=0.05t$},
	legend pos = south east,
	legend style={nodes={scale=0.7, transform shape}},
	xticklabel style = {/pgf/number format/precision=3,
		/pgf/number format/fixed},
	]
	\addplot
	[
	red,
	thin,
	only marks,
	mark=x,
	]
	table
	[
	y expr = \thisrowno{1},
	x expr = \thisrowno{0},
	]
	{Data/sc_orp_over_tp_Pr_1_Phr_1_L_160_n_1_U_-10.0_V_0.05_chi_200.dat};
	\addplot
	[
	red,
	thin,
	only marks,
	mark=o,
	]
	table
	[
	y expr = \thisrowno{1},
	x expr = \thisrowno{0},
	]
	{Data/sc_orp_over_tp_Pr_1_Phr_1_L_160_n_1_U_-10.0_V_0.05_chi_100.dat};
	\addplot
	[
	red,
	thin,
	only marks,
	mark=|,
	]
	table
	[
	y expr = \thisrowno{1},
	x expr = \thisrowno{0},
	]
	{Data/sc_orp_over_tp_Pr_1_Phr_1_L_96_n_1_U_-10.0_V_0.05_chi_200.dat};
	\addplot
	[
	red,
	thin,
	only marks,
	mark=-,
	]
	table
	[
	y expr = \thisrowno{1},
	x expr = \thisrowno{0},
	]
	{Data/sc_orp_over_tp_Pr_1_Phr_1_L_96_n_1_U_-10.0_V_0.05_chi_100.dat};
	\nextgroupplot
	[
	width=0.33\textwidth,
	xlabel = {$t_\perp/t$},
	ylabel = {$|\braket{\hat n_i}_{n}-\braket{\hat n_i}_{n-1}|$},
	legend pos = south east,
	legend style={nodes={scale=0.7, transform shape}},
	xticklabel style = {/pgf/number format/precision=3,
		/pgf/number format/fixed},
	]
	\addplot
	[
	red,
	thin,
	only marks,
	mark=x,
	]
	table
	[
	y expr = \thisrowno{1},
	x expr = \thisrowno{0},
	]
	{Data/cdw_orp_over_tp_Pr_1_Phr_1_L_160_n_1_U_-10.0_V_-0.05_chi_200.dat};
	\addplot
	[
	red,
	thin,
	only marks,
	mark=o,
	]
	table
	[
	y expr = \thisrowno{1},
	x expr = \thisrowno{0},
	]
	{Data/cdw_orp_over_tp_Pr_1_Phr_1_L_160_n_1_U_-10.0_V_-0.05_chi_100.dat};
	\addplot
	[
	red,
	thin,
	only marks,
	mark=|,
	]
	table
	[
	y expr = \thisrowno{1},
	x expr = \thisrowno{0},
	]
	{Data/cdw_orp_over_tp_Pr_1_Phr_1_L_96_n_1_U_-10.0_V_-0.05_chi_200.dat};
	\addplot
	[
	red,
	thin,
	only marks,
	mark=-,
	]
	table
	[
	y expr = \thisrowno{1},
	x expr = \thisrowno{0},
	]
	{Data/cdw_orp_over_tp_Pr_1_Phr_1_L_96_n_1_U_-10.0_V_-0.05_chi_100.dat};
	\nextgroupplot
	[
	width=0.33\textwidth,
	xlabel = {$t_\perp/t$},
	legend pos = south east,
	legend style={nodes={scale=0.7, transform shape}},
	xticklabel style = {/pgf/number format/precision=3,
		/pgf/number format/fixed},
	yticklabel style = {/pgf/number format/precision=3,
		/pgf/number format/fixed},
	scaled y ticks = false,
	]
	\addplot
	[
	red,
	thin,
	only marks,
	mark=x,
	]
	table
	[
	y expr = \thisrowno{1},
	x expr = \thisrowno{0},
	]
	{Data/cdw_orp_over_tp_Pr_1_Phr_1_L_160_n_1_U_-10.0_V_0.0_chi_200.dat};
	\addplot
	[
	red,
	thin,
	only marks,
	mark=o,
	]
	table
	[
	y expr = \thisrowno{1},
	x expr = \thisrowno{0},
	]
	{Data/cdw_orp_over_tp_Pr_1_Phr_1_L_160_n_1_U_-10.0_V_0.0_chi_100.dat};
	\addplot
	[
	red,
	thin,
	only marks,
	mark=|,
	]
	table
	[
	y expr = \thisrowno{1},
	x expr = \thisrowno{0},
	]
	{Data/cdw_orp_over_tp_Pr_1_Phr_1_L_96_n_1_U_-10.0_V_0.0_chi_200.dat};
	\addplot
	[
	red,
	thin,
	only marks,
	mark=-,
	]
	table
	[
	y expr = \thisrowno{1},
	x expr = \thisrowno{0},
	]
	{Data/cdw_orp_over_tp_Pr_1_Phr_1_L_96_n_1_U_-10.0_V_0.0_chi_100.dat};
	\nextgroupplot
	[
	width=0.33\textwidth,
	xlabel = {$t_\perp/t$},
	legend pos = south east,
	legend style={nodes={scale=0.7, transform shape}},
	xticklabel style = {/pgf/number format/precision=3,
		/pgf/number format/fixed},
	]
	\addplot
	[
	red,
	thin,
	only marks,
	mark=x,
	]
	table
	[
	y expr = \thisrowno{1},
	x expr = \thisrowno{0},
	]
	{Data/cdw_orp_over_tp_Pr_1_Phr_1_L_160_n_1_U_-10.0_V_0.05_chi_200.dat};
	\addplot
	[
	red,
	thin,
	only marks,
	mark=o,
	]
	table
	[
	y expr = \thisrowno{1},
	x expr = \thisrowno{0},
	]
	{Data/cdw_orp_over_tp_Pr_1_Phr_1_L_160_n_1_U_-10.0_V_0.05_chi_100.dat};
	\addplot
	[
	red,
	thin,
	only marks,
	mark=|,
	]
	table
	[
	y expr = \thisrowno{1},
	x expr = \thisrowno{0},
	]
	{Data/cdw_orp_over_tp_Pr_1_Phr_1_L_96_n_1_U_-10.0_V_0.05_chi_200.dat};
	\addplot
	[
	red,
	thin,
	only marks,
	mark=-,
	]
	table
	[
	y expr = \thisrowno{1},
	x expr = \thisrowno{0},
	]
	{Data/cdw_orp_over_tp_Pr_1_Phr_1_L_96_n_1_U_-10.0_V_0.05_chi_100.dat};
	\end{groupplot}
	\end{tikzpicture}
	\caption{Comparison of measurements with $U=-10t$ and $n=1$ at $L\in[128,160]$ and bond dimension $\chi\in[100,200]$ with various $t_\perp$ and $V$.}
	\label{Fig::Strong_attract_precision}
\end{figure*}
\section{Initial values for self-consistent iteration\label{App::Initial}}
In order to establish the symmetry between \gls{CDW} and \gls{SC} phases for $n=1$ care must be taken to what initial values are chosen for the \gls{MF} amplitudes. In order to produce coexistence we choose
\begin{align}
	\alpha^{(\mathrm{init})}_{i,k} =
	\begin{cases}
	\alpha_{i-k} & \mbox{if } i=k \\
	0 & \mbox{otherwise }
	\end{cases}\;,
\end{align}
which represents a strongly local pairing constant throughout the system. This drives towards a \gls{SC} phase. Conversely, since coexistence occurs at $n=1$ the periodicity of a \gls{CDW} phase is $2$, i.e., the density repeats every other site. Hence, the \gls{CDW} \gls{MF} is initialized with
\begin{equation}
	\mu^{(\mathrm{init})}_{i\sigma} = 
	\begin{cases}
	\mu_{\sigma} & \mbox{if i even} \\
	-\mu_{\sigma} & \mbox{if i odd}
	\end{cases}\;,
\end{equation}
with the additional condition that $\mu_\uparrow=\mu_\downarrow$. This choice drives towards a \gls{CDW} phase with period 2. Lastly the particle-hole are not deemed as leading drivers toward any order and are initialized at zero:
\begin{equation}
	\beta^{(\mathrm{init})}_{i,r,\sigma} = 0\;.
\end{equation}

\subsection{Doped CDW}
When doping away from $n=1$ the \gls{SC} phase should remain equally simple: Constant local pairing throughout the system. Conversely, the \gls{CDW} phase can longer be argued to have a simple period of 2 in the density profile. In fact, theory would expect each \gls{1D} system to obtain a period of $2/n$~\cite{Giamarchi2003}. However, light doping causes density in a discrete system to repeat only at a large number of sites.

One method, used in this work, to obtain reasonable initial values for \gls{CDW} in a doped system is given by letting $\alpha_{i,k}=\beta_{i,r,\sigma}=0$ at all iterations. With this restriction no pairing is possible and only the \gls{CDW} \gls{MF} is kept in the solution leaving no other ordering channel than a \gls{CDW}. The resulting solution is kept as initial for a competition between \gls{CDW} and \gls{SC} phases.

\section{Hermiticity of two-period solution\label{App::Hermiticity}}
One potential concern for the two-periodic behavior is that the Hamiltonian strongly disrespects translational symmetry. Due to this property care must be taken that qualities such as hermiticity are preserved: Tunneling out of a high-density site into a low-density one could possibly have different amplitude than its conjugate. Such an observation is relevant for the two-periodic behavior associated with the occurrence of \gls{CDW}. In order to resolve this, consider the ground state of one of the two-period solutions
\begin{equation}
\ket{\psi} = \ket{\phi} + \ket{\mathrm{CDW}}\;,
\end{equation}
where the \gls{CDW} part contains all two-periodic behavior and $\ket{\phi}$ single period effects. For unit filling the density is large at every other site and almost zero at the others. It makes intuitive sense that a tunneling event out of a high-density region is more probable than into one yet such events are conjugate.

Using $\ket{\psi}$ the measurement becomes
\begin{flalign}
\braket{\psi|\hat c^{\dagger}_i \hat c^{\nodagger}_j|\psi} = &\braket{\phi|\hat c^{\dagger}_i \hat c^{\nodagger}_j|\phi} + \braket{\mathrm{CDW}|\hat c^{\dagger}_i \hat c^{\nodagger}_j|\mathrm{CDW}}\nonumber\\& + \braket{\phi|\hat c^{\dagger}_i \hat c^{\nodagger}_j|\mathrm{CDW}} + \braket{\mathrm{CDW}|\hat c^{\dagger}_i \hat c^{\nodagger}_j|\phi}\;,
\end{flalign}
and the conjugate process with $i\xleftrightarrow{}j$. Performing the application and writing down amplitudes yields
\begin{flalign}
\braket{\psi|\hat c^{\dagger}_i \hat c^{\nodagger}_j|\psi} &= \alpha_1\underbrace{\braket{\phi|\phi'}}_{\beta_1} + \alpha_2\underbrace{\braket{\mathrm{CDW}|\mathrm{CDW}'}}_{\beta_2}\nonumber \\ &+ \alpha_3\underbrace{\braket{\phi|\mathrm{CDW}'}}_{\beta_3} + \alpha_4\underbrace{\braket{\mathrm{CDW}|\phi'}}_{\beta_4}\;,
\end{flalign}
where $\alpha_i$ indicate the amplitudes of the surviving states after application of $\hat c^{\dagger}_i \hat c^{\nodagger}_j$. The $\beta_i$ then indicate the amplitudes of the states which contain finite overlap with the primed states. Importantly, the primed states contain a single particle moved from $j$ to $i$. Thus, any finite $\beta_i$ result from having states in the primed superpositions which are already inside the unprimed ones. However, if those states are already inside the unprimed superpositions then the conjugate operation $\hat c^{\dagger}_j \hat c^{\nodagger}_i$ generates the same states as those spared by application of $\hat c^{\dagger}_i \hat c^{\nodagger}_j$. Hence,
\begin{flalign}
\braket{\psi|\hat c^{\dagger}_j \hat c^{\nodagger}_i|\psi} =& \beta_1^*\underbrace{\braket{\phi|\phi''}}_{\alpha_1^*} + \beta_2^*\underbrace{\braket{\mathrm{CDW}|\mathrm{CDW}''}}_{\alpha_2^*}\nonumber \\ &+ \beta_3^*\underbrace{\braket{\phi|\mathrm{CDW}''}}_{\alpha_3^*} + \beta_4^*\underbrace{\braket{\mathrm{CDW}|\phi''}}_{\alpha_4^*}\;,
\end{flalign}
which indicates that $\braket{\hat c^{\dagger}_i \hat c^{\nodagger}_j}=\braket{\hat c^{\dagger}_j \hat c^{\nodagger}_i}$ if the amplitudes are real.

\section{Precision of simulation\label{App::Precision}}
\begin{figure*}[t!]
	\tikzsetnextfilename{PrecisionQualityCheck2U}
	\tikzset{external/export next=true}
	\centering
	\begin{tikzpicture}
	\begin{groupplot}
	[group style = {group size=3 by 2}]
	\nextgroupplot
	[
	width=0.33\textwidth,
	title = {$n=0.875$},
	ylabel = {$\braket{\hat c_{i,\uparrow} \hat c_{i,\downarrow}}$},
	legend pos = north east,
	legend style={nodes={scale=0.7, transform shape}},
	xticklabel style = {/pgf/number format/precision=3,
		/pgf/number format/fixed},
	yticklabel style = {/pgf/number format/precision=3,
		/pgf/number format/fixed},
	scaled x ticks = false,
	scaled y ticks = false,
	]
	\addplot
	[
	red,
	thin,
	only marks,
	mark=x,
	]
	table
	[
	y expr = \thisrowno{1},
	x expr = \thisrowno{0},
	]
	{Data/sc_orp_over_V_Pr_5_Phr_5_tp_0.05_L_192_n_0.875_U_-2.0_chi_600.dat};
	\addlegendentry{$L=192,\chi=600$}
	\addplot
	[
	red,
	thin,
	only marks,
	mark=o,
	]
	table
	[
	y expr = \thisrowno{1},
	x expr = \thisrowno{0},
	]
	{Data/sc_orp_over_V_Pr_5_Phr_5_tp_0.05_L_192_n_0.875_U_-2.0_chi_400.dat};
	\addlegendentry{$L=192,\chi=400$}
	\addplot
	[
	red,
	thin,
	only marks,
	mark=|,
	]
	table
	[
	y expr = \thisrowno{1},
	x expr = \thisrowno{0},
	]
	{Data/sc_orp_over_V_Pr_5_Phr_5_tp_0.05_L_128_n_0.875_U_-2.0_chi_600.dat};
	\addlegendentry{$L=128,\chi=600$}
	\addplot
	[
	red,
	thin,
	only marks,
	mark=-,
	]
	table
	[
	y expr = \thisrowno{1},
	x expr = \thisrowno{0},
	]
	{Data/sc_orp_over_V_Pr_5_Phr_5_tp_0.05_L_128_n_0.875_U_-2.0_chi_400.dat};
	\addlegendentry{$L=128,\chi=400$}
	\nextgroupplot
	[
	width=0.33\textwidth,
	title = {$n=0.9375$},
	legend pos = south east,
	legend style={nodes={scale=0.7, transform shape}},
	xticklabel style = {/pgf/number format/precision=3,
		/pgf/number format/fixed},
	yticklabel style = {/pgf/number format/precision=3,
		/pgf/number format/fixed},
	scaled y ticks = false,
	scaled x ticks = false,
	]
	\addplot
	[
	red,
	thin,
	only marks,
	mark=x,
	]
	table
	[
	y expr = \thisrowno{1},
	x expr = \thisrowno{0},
	]
	{Data/sc_orp_over_V_Pr_5_Phr_5_tp_0.05_L_192_n_0.9375_U_-2.0_chi_600.dat};
	\addplot
	[
	red,
	thin,
	only marks,
	mark=o,
	]
	table
	[
	y expr = \thisrowno{1},
	x expr = \thisrowno{0},
	]
	{Data/sc_orp_over_V_Pr_5_Phr_5_tp_0.05_L_192_n_0.9375_U_-2.0_chi_400.dat};
	\addplot
	[
	red,
	thin,
	only marks,
	mark=|,
	]
	table
	[
	y expr = \thisrowno{1},
	x expr = \thisrowno{0},
	]
	{Data/sc_orp_over_V_Pr_5_Phr_5_tp_0.05_L_128_n_0.9375_U_-2.0_chi_600.dat};
	\addplot
	[
	red,
	thin,
	only marks,
	mark=-,
	]
	table
	[
	y expr = \thisrowno{1},
	x expr = \thisrowno{0},
	]
	{Data/sc_orp_over_V_Pr_5_Phr_5_tp_0.05_L_128_n_0.9375_U_-2.0_chi_400.dat};
	\nextgroupplot
	[
	width=0.33\textwidth,
	title = {$n=1$},
	legend pos = south east,
	legend style={nodes={scale=0.7, transform shape}},
	xticklabel style = {/pgf/number format/precision=3,
		/pgf/number format/fixed},
	yticklabel style = {/pgf/number format/precision=3,
		/pgf/number format/fixed},
	scaled y ticks = false,
	scaled x  ticks = false,
	]
	\addplot
	[
	red,
	thin,
	only marks,
	mark=x,
	]
	table
	[
	y expr = \thisrowno{1},
	x expr = \thisrowno{0},
	]
	{Data/sc_orp_over_V_Pr_5_Phr_5_tp_0.05_L_192_n_1_U_-2.0_chi_600.dat};
	\addplot
	[
	red,
	thin,
	only marks,
	mark=o,
	]
	table
	[
	y expr = \thisrowno{1},
	x expr = \thisrowno{0},
	]
	{Data/sc_orp_over_V_Pr_5_Phr_5_tp_0.05_L_192_n_1_U_-2.0_chi_400.dat};
	\addplot
	[
	red,
	thin,
	only marks,
	mark=|,
	]
	table
	[
	y expr = \thisrowno{1},
	x expr = \thisrowno{0},
	]
	{Data/sc_orp_over_V_Pr_5_Phr_5_tp_0.05_L_128_n_1_U_-2.0_chi_600.dat};
	\addplot
	[
	red,
	thin,
	only marks,
	mark=-,
	]
	table
	[
	y expr = \thisrowno{1},
	x expr = \thisrowno{0},
	]
	{Data/sc_orp_over_V_Pr_5_Phr_5_tp_0.05_L_128_n_1_U_-2.0_chi_400.dat};
	\nextgroupplot
	[
	width=0.33\textwidth,
	xlabel = {$V/t$},
	ylabel = {$|\braket{\hat n_i}_{n}-\braket{\hat n_i}_{n-1}|$},
	legend pos = south east,
	legend style={nodes={scale=0.7, transform shape}},
	xticklabel style = {/pgf/number format/precision=3,
		/pgf/number format/fixed},
	scaled x ticks = false,
	]
	\addplot
	[
	red,
	thin,
	only marks,
	mark=x,
	]
	table
	[
	y expr = \thisrowno{1},
	x expr = \thisrowno{0},
	]
	{Data/cdw_orp_over_V_Pr_5_Phr_5_tp_0.05_L_192_n_0.875_U_-2.0_chi_600.dat};
	\addplot
	[
	red,
	thin,
	only marks,
	mark=o,
	]
	table
	[
	y expr = \thisrowno{1},
	x expr = \thisrowno{0},
	]
	{Data/cdw_orp_over_V_Pr_5_Phr_5_tp_0.05_L_192_n_0.875_U_-2.0_chi_400.dat};
	\addplot
	[
	red,
	thin,
	only marks,
	mark=|,
	]
	table
	[
	y expr = \thisrowno{1},
	x expr = \thisrowno{0},
	]
	{Data/cdw_orp_over_V_Pr_5_Phr_5_tp_0.05_L_128_n_0.875_U_-2.0_chi_600.dat};
	\addplot
	[
	red,
	thin,
	only marks,
	mark=-,
	]
	table
	[
	y expr = \thisrowno{1},
	x expr = \thisrowno{0},
	]
	{Data/cdw_orp_over_V_Pr_5_Phr_5_tp_0.05_L_128_n_0.875_U_-2.0_chi_400.dat};
	\nextgroupplot
	[
	width=0.33\textwidth,
	xlabel = {$V/t$},
	legend pos = south east,
	legend style={nodes={scale=0.7, transform shape}},
	xticklabel style = {/pgf/number format/precision=3,
		/pgf/number format/fixed},
	yticklabel style = {/pgf/number format/precision=3,
		/pgf/number format/fixed},
	scaled x ticks = false
	]
	\addplot
	[
	red,
	thin,
	only marks,
	mark=x,
	]
	table
	[
	y expr = \thisrowno{1},
	x expr = \thisrowno{0},
	]
	{Data/cdw_orp_over_V_Pr_5_Phr_5_tp_0.05_L_192_n_0.9375_U_-2.0_chi_600.dat};
	\addplot
	[
	red,
	thin,
	only marks,
	mark=o,
	]
	table
	[
	y expr = \thisrowno{1},
	x expr = \thisrowno{0},
	]
	{Data/cdw_orp_over_V_Pr_5_Phr_5_tp_0.05_L_192_n_0.9375_U_-2.0_chi_400.dat};
	\addplot
	[
	red,
	thin,
	only marks,
	mark=|,
	]
	table
	[
	y expr = \thisrowno{1},
	x expr = \thisrowno{0},
	]
	{Data/cdw_orp_over_V_Pr_5_Phr_5_tp_0.05_L_128_n_0.9375_U_-2.0_chi_600.dat};
	\addplot
	[
	red,
	thin,
	only marks,
	mark=-,
	]
	table
	[
	y expr = \thisrowno{1},
	x expr = \thisrowno{0},
	]
	{Data/cdw_orp_over_V_Pr_5_Phr_5_tp_0.05_L_128_n_0.9375_U_-2.0_chi_400.dat};
	\nextgroupplot
	[
	width=0.33\textwidth,
	xlabel = {$V/t$},
	legend pos = south east,
	legend style={nodes={scale=0.7, transform shape}},
	xticklabel style = {/pgf/number format/precision=3,
		/pgf/number format/fixed},
	scaled x ticks = false,
	]
	\addplot
	[
	red,
	thin,
	only marks,
	mark=x,
	]
	table
	[
	y expr = \thisrowno{1},
	x expr = \thisrowno{0},
	]
	{Data/cdw_orp_over_V_Pr_5_Phr_5_tp_0.05_L_192_n_1_U_-2.0_chi_600.dat};
	\addplot
	[
	red,
	thin,
	only marks,
	mark=o,
	]
	table
	[
	y expr = \thisrowno{1},
	x expr = \thisrowno{0},
	]
	{Data/cdw_orp_over_V_Pr_5_Phr_5_tp_0.05_L_192_n_1_U_-2.0_chi_400.dat};
	\addplot
	[
	red,
	thin,
	only marks,
	mark=|,
	]
	table
	[
	y expr = \thisrowno{1},
	x expr = \thisrowno{0},
	]
	{Data/cdw_orp_over_V_Pr_5_Phr_5_tp_0.05_L_128_n_1_U_-2.0_chi_600.dat};
	\addplot
	[
	red,
	thin,
	only marks,
	mark=-,
	]
	table
	[
	y expr = \thisrowno{1},
	x expr = \thisrowno{0},
	]
	{Data/cdw_orp_over_V_Pr_5_Phr_5_tp_0.05_L_128_n_1_U_-2.0_chi_400.dat};
	\end{groupplot}
	\end{tikzpicture}
	\caption{Comparison of measurements at $U=-2t$ and $t_\perp=0.05t$ for $L\in[128,192]$ and bond dimension $\chi\in[400,600]$, varying $n$ and $V$.}
	\label{Fig::Weak_attract_precision}
\end{figure*}
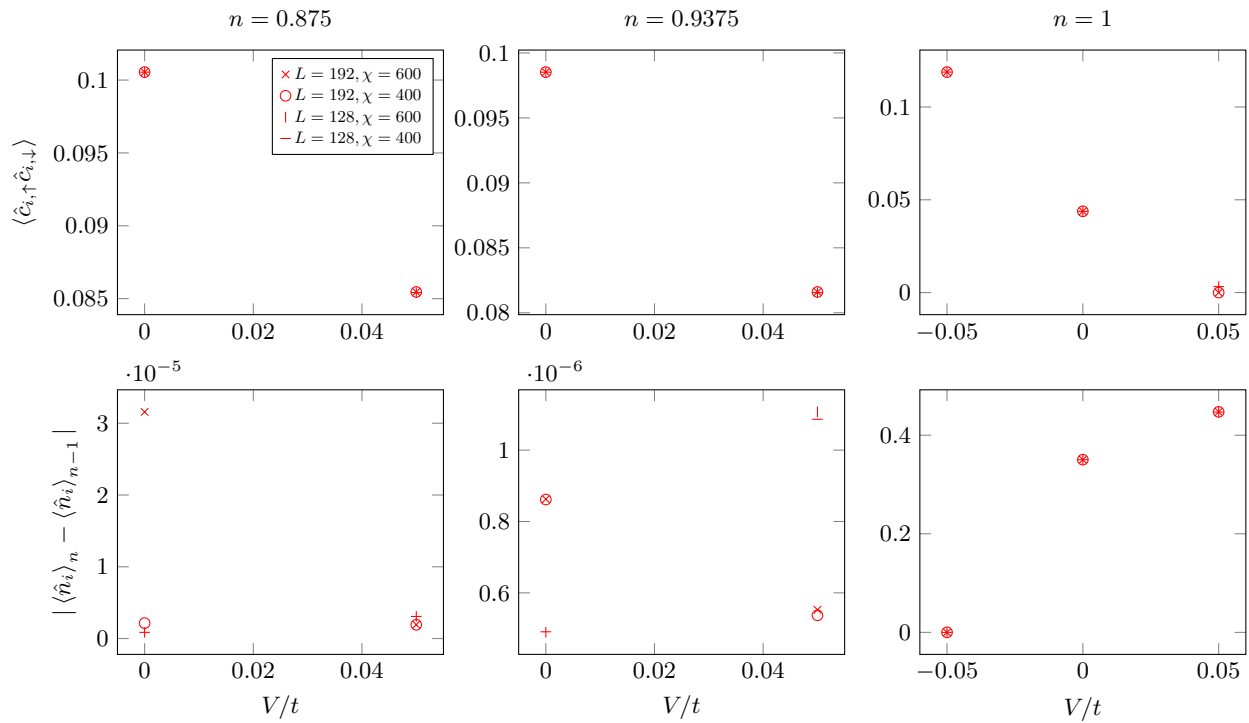

While simulation is done with quasi-exact methods in combination with mean-field theory the former always carry a quantitative error. Since it is not the aim of this paper to precisely compute any property of the system under study the commonly applied linear extrapolation in truncation error is not utilized~\cite{White1993b,Legeza1996}. Instead, a much more qualitative comparison is performed where the precision parameters are varied and we check for notable changes.

It is important to note that errors in the state due to insufficient bond dimension can be more than a question of precision. In fact, if bond dimension is taken low enough qualitative differences can appear as well. An example of such a difference is when an incorrect phase for the ground state is realized due to insufficient bond dimension to capture the true ground state. In order to ward against this issue we show the order of magnitude for truncation errors associated with our simulations in \cref{Tab::trunc_table}~\cite{Schollwock2005}. In general, we find that given enough bond dimension the truncation error goes towards zero.

Additionally, since the \gls{DMRG} routine is run in a finite size manner we also expect finite size effects. Assuming that very large sizes correctly capture the essential physics we compare measurements for different sizes and bond dimension in \cref{Fig::Strong_attract_precision,Fig::Weak_attract_precision}. Notably, we find that neither size nor bond dimension has a large qualitative impact on the measurements.
\begin{table}
	\centering
	\renewcommand{\arraystretch}{1.5}
	\begin{tabular}{||c|c|c|c|c||}
		\hline
		 \backslashbox{$U/t$}{$\chi$}& $100$ & $200$ & $400$ & $600$ \\\hline
		 $-10.0$ & $10^{-9}$ & $10^{-13}$ & -- & -- \\\hline
		 $-2.0$ & $10^{-4}$ & $10^{-5}$ & $10^{-7}$ & $10^{-8}$ \\\hline
	\end{tabular}
	\caption{Table showing the order of magnitude of truncation error $\epsilon_\psi$ for the various bond dimension $\chi$ used.}
	\label{Tab::trunc_table}
\end{table}
\bibliography{gunnar,adrian}

\end{document}